\documentclass[aps,pra,twocolumn]{revtex4-2}

\pdfoutput=1

\usepackage{amsfonts,amsmath,amssymb}

\usepackage{graphicx}
\graphicspath{{graphics/}} 
\usepackage{xcolor}

\usepackage{lmodern} 
\usepackage[T1]{fontenc}
\usepackage[utf8]{inputenc} 
\usepackage{amssymb,amsmath} 
\usepackage{textcomp}
\usepackage{mathrsfs}
\usepackage{braket}
\usepackage{mathtools}

\newcommand{\de}{\partial}
\renewcommand{\L}{\mathcal{L}}
\newcommand{\e}{\epsilon}

\newcommand{\p}{\phi}
\renewcommand{\b}{\beta}
\newcommand{\m}{\mu}

\newcommand{\n}{\nu}
\newcommand{\del}{\nabla}

\renewcommand{\d}{\mathrm{d}}

\newcommand{\vp}{\varphi}
\newcommand{\T}{\Theta}
\newcommand{\w}{\omega}

\newcommand{\vc}[1]{\mathbf{#1}}

\newcommand{\h}{\hbar}
\newcommand{\Tr}{\mathcal{T}}
\newcommand{\R}{\mathcal{R}}

\newcommand{\abs}[1]{\left| #1 \right|}

\newcommand{\tld}[1]{\widetilde{#1}}

\begin{document}

\title{Understanding superradiant phenomena with synthetic vector potentials in atomic Bose--Einstein condensates}

\author{Luca Giacomelli}
\email[]{luca.giacomelli-1@unitn.it}
\author{Iacopo Carusotto}
\email[]{iacopo.carusotto@unitn.it}
\affiliation{INO-CNR BEC Center and Dipartimento di Fisica, Universit\`a di Trento, via Sommarive 14, I-38050 Povo, Trento, Italy}

\begin{abstract}
	We theoretically investigate superradiance effects in quantum field theories in curved space-times by proposing an analogue model based on Bose--Einstein condensates subject to a synthetic vector potential. The breaking of the irrotationality constraint of superfluids allows to study superradiance in simple planar geometries and obtain intuitive insight in the amplified scattering processes at ergosurfaces. When boundary conditions are modified allowing for reflections, dynamical instabilities are found, similar to the ones of ergoregions in rotating space-times. Their stabilization by horizons in black hole geometries is discussed. 
	All these phenomena are reinterpreted through an exact mapping with the physics of one-dimensional relativistic charged scalar fields in electrostatic potentials. 
	Our study provides a deeper understanding on the basic mechanisms of superradiance: by disentangling the different ingredients at play, it shines light on some misconceptions on the role of dissipation and horizons and on the competition between superradiant scattering and instabilities. 
\end{abstract}

\maketitle

\section{Introduction}

The term {\em superradiance} is typically used in the context of quantum field theories and gravitational physics to indicate generic processes where some radiation gets amplified upon scattering onto some object~\cite{brito2015superradiance}. As such, it is a very general phenomenon appearing in many different physical systems: celebrated examples of it are the bosonic Klein paradox for charged fields incident on a step-like electrostatic potential, the Zel'dovich amplification of electromagnetic waves from a fast spinning absorbing body, hydrodynamic wave amplification at spatial discontinuities of the flow profile and, finally, the superradiant scattering of scalar field, electromagnetic or gravitational waves from rotating black holes. These amplification processes are often associated to different instabilities mechanisms that are important in the search of physics beyond the standard model.

Superradiance in the gravitational context can be directly translated to condensed matter systems through the so-called gravitational analogy that maps the propagation of a scalar field in a curved space-time onto the one of sound in a non-uniformly moving fluid~\cite{barcelo2011analogue}. This mapping is possible because superradiance only relies on the kinematics of the fields in a fixed background and not on the gravitational dynamics. In this framework, it has been extensively theoretically studied in vortex configurations that reproduce the essential features of rotating black holes \cite{basak2003superresonance, berti2004quasinormal, federici2006superradiance, richartz2009generalized, richartz2015rotating, cardoso2016detecting} and has been recently observed in such a setup using gravity waves on water \cite{torres2017rotational}.

In spite of these impressive advances, there are still a number of intriguing open points in our understanding of basic superradiance phenomena. The circular geometry of rotating systems and their limited tunability makes it difficult to disentangle the different microscopic mechanisms at play and build an intuitive picture of the overall process. The goal of the present work is to propose a new concept of analog model that allows a local tuning of the velocity profile and, thus, a wider flexibility in the design of the analog space-time to be studied. In this way, we obtain a comprehensive understanding of the basic superradiant phenomena and of the related dynamical instability mechanisms.

The key idea of our work is to use an atomic Bose--Einstein condensate (BEC)~\cite{pitaevskii2016bose} subject to a so-called synthetic vector potential~\cite{dalibard2011colloquium}. Several strategies to this purpose have been demonstrated in the recent years using combinations of static electromagnetic fields and microwave and/or optical Raman beams~\cite{lin2009synthetic, spielman2009raman, lin2009bose}. As a result, neutral atoms end up experiencing a minimal coupling to a vector potential that is formally analogous to the electrostatic vector potential acting on electrically charged objects and is responsible for all sorts of magnetic effects~\cite{RMP_Dali_2019}. Even though we will restrict here to the case of atomic systems, it is worth reminding that similar routes can also be explored for analog models based on optical systems~\cite{nguyen2015,Solnyshkov_2011,Gerace_2012} where synthetic magnetic fields for neutral photons are presently the subject of intense investigations~\cite{RMP_2019}.

The basic effect of a synthetic vector potential is to change the relation between the wavevector of the associated matter wave to the physical velocity of the atoms. This allows to break the irrotationality constraint of superfluid flows and, thus, widens the range of spatial flow profiles that can be generated and used as analog models. In this work, we exploit this idea to propose a setup in which superradiance occurs in a simple and tunable geometry displaying a jump in the transverse velocity. Here, superradiance can be understood in terms of the scattering of a plane wave on a single planar interface playing the role of the analogue of a black hole \textit{ergosurface}. 
This approach takes inspiration from recent investigations of analogue Hawking radiation from analog black holes, where the horizon was created by juxtaposing two uniform regions, a subsonic one and a supersonic one \cite{balbinot2008nonlocal,carusotto2008numerical,recati2009bogoliubov}.

Further interest for these simple configurations is provided by the possibility of an exact mapping via dimensional reduction onto the one-dimensional problem of a charged and massive relativistic scalar field incident coupled to an electrostatic potential. This offers further evidence that the superradiance of the bosonic Klein paradox and black hole superradiance are essentially two manifestations of the same phenomenon \cite{fulling1989aspects} and provides a novel route for the quantum simulation of charged massive quantum fields in complex electrostatic potential landscapes.

By changing the boundary conditions, e.g. by introducing reflection on one of the two sides of the ergosurface, transitions from a stable superradiant amplification to dynamical instability behaviours can occur. These mechanisms are analogous to the instabilities predicted for rotating black holes when they are placed in a reflecting box or an asymptotically anti-de Sitter space-time, or for horizonless rotating stars~\cite{brito2015superradiance}. Also
in the electrostatic mapping similar instabilities occur with box-shaped electrostatic potentials via the so-called Schiff--Snyder--Weinberg effect \cite{schiff1940existence, fulling1989aspects}. One of the objectives of our study is to use the analog model to shine light on all these intriguing instability phenomena and build a unified and intuitive picture of them all.

As a final step, we introduce an horizon into the model and we investigate how this affects the dynamical instabilities. As a remarkable result, we find that horizons do not in general prevent the development of instabilities in the ergoregion of analog black holes, as instead is usually the case in general-relativistic black holes. For suitable choices of parameters, the waves travelling towards the horizon can in fact undergo a sizable reflection leading to a dynamical instability.

Putting these questions in an even wider perspective, we revisit widespread statements in the literature about the actual need of dissipating elements for the observation of superradiance effects. While dynamical stability of the system requires that the superradiant amplified field is efficiently evacuated outside of the system or dissipated, we confirm the result in~\cite{vicente2018penrose} that absence of dissipation does not prevent a wavepacket from being efficiently amplified by superradiant processes on a shorter time scale.

Another common thread that extends throughout the whole work is to consider the impact of the super-luminal dispersion of the Bogoliubov sound in BECs onto the different superradiant effects under investigation. As a general result, in agreement with our previous results for multiply quantized vortices~\cite{giacomelli2019ergoregion}, we find that amplified reflections and instabilities are typically not affected by the dispersion at low momenta and preserve all their qualitative features, while they are strongly suppressed at large momenta. 

While our setup is not expected to quantitatively reproduce the physics of specific general-relativistic systems, it offers a {\em toy model} in which the fundamental mechanisms of superradiance are reproduced in the simplest possible setting and each of the different elements at play can be addressed individually. This provides a conceptual framework that includes astrophysical black holes as particular cases of a more general effect and, thus, can serve as a guide in the study of new configurations.

The structure of this work is the following.
In Section \ref{sec:metric} we show how a synthetic vector potential for a BEC modifies the acoustic metric of the corresponding curved space-time and how this possibility can be exploited to increase the range of curved space-time to be investigated in analog models. 
In Section \ref{sec:transverseflow} we study the Klein--Gordon equation for a \textit{minimal} acoustic metric showing superradiance. The kinematic structure of the process is characterized in Section \ref{sec:superradiant_scatt}. In Section \ref{sec:electrostatic} we exploit the exact mapping with an electrostatic problem to derive the condition for amplified scattering. In Section \ref{sec:dispersion} we comment on the modifications due to dispersion and in Section \ref{sec:gpe} we verify the occurrence of superradiant scattering with a numerical study of the full dynamics of the condensate. Then in Section \ref{sec:instabilities} we study the occurrence of dynamical instabilities when the boundary conditions are changed and reflections are introduced. These instabilities are illustrated by means of numerical simulations in Sec.\ref{sec:instab_nume}, in terms of the Bogoliubov excitation spectra in Sec.\ref{sec:instab_Bogo} and via mode-matching techniques in Sec.\ref{sec:modematch}. They are then linked to the Schiff--Snyder--Weinberg effect for massive charged fields incident on box-shaped electrostatic potentials and to the instabilities of astrophysical objects in Sec.\ref{sec:discuss}. In Section \ref{sec:horizon} we further extend the model by including an acoustic horizon and investigating how this latter affects superradiant  phenomena: in Sec.\ref{sec:horizon_scatt} we discuss scattering features at horizons, in Sec.\ref{sec:toy_bh} we describe our toy-model setup, and in Sec.\ref{sec:horizon_dyn} we investigate dynamical instabilities in the presence of an horizon. Conclusions and future perspectives are finally sketched in Sec.\ref{sec:conclusions}. Additional material on the relation between the Bogoliubov problem and the hydrodynamic approximation encoded in the Klein-Gordon equation is given in the Appendix.

\section{Acoustic metric in the presence of a synthetic vector potential}\label{sec:metric}

At the mean-field level, the dynamics of a dilute Bose-Einstein condensate of weakly interacting bosons can be described in terms of a complex scalar field $\Psi(\vc{r},t)$, corresponding to the order parameter of the condensation phase transition and obeying the famous Gross--Pitaevskii equation (GPE) \cite{pitaevskii2016bose}. For the sake of simplicity, in this work we consider a two-dimensional condensate, where one dimension is frozen by a tight confinement and the dynamics can be described by a two-dimensional GPE. Generalization to the three-dimensional case would make the discussion more cumbersome but would not involve any additional conceptual difficulty.

In the last decades, strong efforts have been devoted to the design of combinations of optical and/or microwave and/or magnetostatic fields that result in a modification of the effective dispersion relation of the atoms and, in particular, in a shift of the position of its minimum. In mathematical terms, this can be described by including additional terms to the GPE in the form of a vector potential $\vc{A}(\vc{r},t)$ minimally coupled to the atomic momentum, the so-called {\em synthetic vector potential}~\cite{dalibard2011colloquium,leblanc_spielman_2017}~\footnote{Note that depending on the microscopic scheme used to generate the synthetic vector potential, additional correction terms may arise~\cite{dalibard2011colloquium}. For the sake of simplicity and generality, in the following we will work under the assumption that they are negligible.}. Putting all terms together, the GPE then reads
\begin{equation}\label{eq:gpe_gauge}
	i\h \de_t \Psi = \left[\frac{(-i\h\del-\vc{A}(\vc{r},t))^2}{2M}+V(\vc{r})+g\abs{\Psi}^2\right]\Psi\,,
\end{equation}
where $V(\vc{r})$ is the external trapping potential, $g$ is the interaction constant and $M$ is the bare atomic mass. In analogy to usual magnetism, the curl of the vector potential provides the {\em synthetic magnetic field} $\vc{B}$ experienced by the neutral atoms. Its time dependence contributes to the {\em synthetic electric field}. In what follows, we will focus our attention on static vector potentials with complex spatial profiles $\vc{A}(\vc{r})$ giving spatially inhomogeneous synthetic magnetic fields but no synthetic electric field.

The core idea of analogue models of gravity is that propagation of sound in a flowing fluid can be described in a geometric way in terms of the propagation of the perturbations of a scalar field on a curved space-time with a non-trivial \textit{acoustic metric} determined by the density and current profile of the underlying fluid~\cite{barcelo2011analogue}. The goal of this section is to show how the presence of the synthetic vector potential modifies the acoustic metric and discuss the new possibilities that this feature allows.

To this end we follow the usual strategy and rewrite the GPE \eqref{eq:gpe_gauge} in the superfluid hydrodynamic form in terms of the modulus and the phase of the order parameter $\Psi=\sqrt{n}\,e^{i\Theta}$:
\begin{subequations}
\begin{align}
	\displaystyle
		&\de_t n = -\del\cdot\left[n\frac{\left(\h\del\T - \vc{A}\right)}{M}\right] \\
		\displaystyle
		&\h\de_t \T + \frac{\left(\h\del\T - \vc{A}\right)^2}{2M}+gn+V-\frac{\h^2}{2M}\frac{\del^2\sqrt{n}}{\sqrt{n}}=0. \label{eq:Euler}
\end{align}
\end{subequations}
Except for the last \textit{quantum pressure} term in the second equation, these equations recover the usual continuity and Euler equations of a perfect fluid.

The effect of the synthetic field is visible in the expression of the velocity field in terms of the condensate phase,
\begin{equation}
	\vc{v}=\frac{\h\del\T - \vc{A}}{M}=\vc{v}_{\rm can}-\frac{\vc{A}}{M}.
\end{equation}
The distinction between the  \textit{canonical velocity} $\vc{v}_{\rm can}=\,\del\T$ and the \textit{physical velocity} $\vc{v}=\vc{v}_{\rm can}-\vc{A}/M$ in the presence of a vector potential has the crucial consequence~\cite{leblanc_spielman_2017} that the physical velocity field $\vc{v}(\vc{r})$ appearing in the hydrodynamic equations is no longer constrained to be irrotational as it occurs in textbook superfluid hydrodynamics. This is the key additional element that synthetic vector potentials introduce in the world of analog models and will be at the heart of all developments in this work.

By linearizing the hydrodynamic equations around some stationary background configuration $\Psi_0(\vc{r})=\sqrt{n_0(\vc{r})}\,e^{i\Theta_0(\vc{r})}$ so that the total density and phase are $n=n_0+n_1$ and $\T=\T_0+\T_1$ one obtains the Bogoliubov-de Gennes equations
\begin{equation}\label{eq:bdg}
\begin{aligned}
		\displaystyle &\de_t n_1 +\del\cdot\left[n_1\vc{v}+n_0 \frac{\h}{M}\del\T_1\right]=0
		\\
		\displaystyle &\h\de_t \T_1 +\h\vc{v}\cdot\del\T_1+gn_1 - \frac{\h^2}{4M}\frac{1}{n}\del\cdot\left[n\del\left(\frac{n_1}{n}\right)\right]=0.
\end{aligned}
\end{equation}
In the so-called hydrodynamic limit where the density and phase profiles vary over distances much larger than the healing length of the condensate $\xi=\sqrt{\h^2/Mgn_0}$, we can safely neglect the quantum pressure term in \eqref{eq:Euler}. As in the standard case~\cite{barcelo2011analogue} of a perfect irrotational fluid in the absence of magnetic effects, the motion equation for $\Theta_1(\vc{r},t)$ can then be recast in the form of a Klein--Gordon equation for a massless scalar field 
\begin{equation}\label{eq:kleingordon-cs}
	\frac{1}{\sqrt{-g}}\de_\mu\left(\sqrt{-g}g^{\mu\nu}\de_\nu\T_1\right)=0
\end{equation}
in a curved spacetime of acoustic metric
\begin{equation}
	g_{\m\n}= \frac{n_0}{c_s}
	\begin{bmatrix}
		-\left[c_s^2-\vc{v}^2\right] & -\vc{v}^T
		\\
		\\
		-\vc{v} & I
	\end{bmatrix},
\end{equation}
where $c_s=\sqrt{gn_0/M}$ is the local speed of sound, $I$ is the identity matrix, and $\vc{v}$ is the local physical velocity. 

As a key result of this section, the possibility of breaking the irrotationality constraint on the physical velocity field $\vc{v}$ by means of the synthetic vector potential dramatically expands the range of space-times that can be implemented in analog models and gives a novel degree of freedom in engineering configurations for the study of analogue gravity effects. In the next sections we will exploit this possibility to investigate new aspects of superradiance phenomena.

\section{Superradiance from an isolated planar ergosurface}\label{sec:transverseflow}

Inspired by analogous phenomena for rotating black holes in gravitational physics, superradiance effects have been widely investigated in the context of analog models, in particular for velocity fields in the form of a vortex with a drain. In this flow geometry, the acoustic metric recovers in fact the main properties of a rotating black hole, i.e. an horizon and an ergoregion \cite{brito2015superradiance,barcelo2011analogue}: the horizon corresponds to the locus of points where the radial velocity crosses the speed of sound, while the ergoregion (delimited by the \textit{ergosurface}) is the spatial region where the overall velocity is supersonic and negative energy modes exist. In such configurations, superradiance is visible as the amplified reflection of a wavepacket scattering on the vortex core~\cite{basak2003superresonance, berti2004quasinormal, federici2006superradiance, richartz2015rotating, cardoso2016detecting} and in this form it has been experimentally observed using surface waves on water~\cite{torres2017rotational}. 

Superradiance crucially relies on the presence of an ergoregion, where negative energy excitations are available and can compensate for the extra positive energy of the amplified scattered waves. In standard treatments, it is often argued that a horizon (providing an open boundary condition) or some other kind of dissipation are an essential ingredient to observe superradiant scattering~\cite{richartz2009generalized}. This point was recently clarified in \cite{vicente2018penrose}, where it was argued that horizons are not an essential ingredient for superradiance. In what follows, we will exploit the wider flexibility offered by the synthetic vector potential to design new configurations that allow to disentangle the role of the different elements at play.

\begin{figure}[t]
  \centering
  \includegraphics[width=\columnwidth]{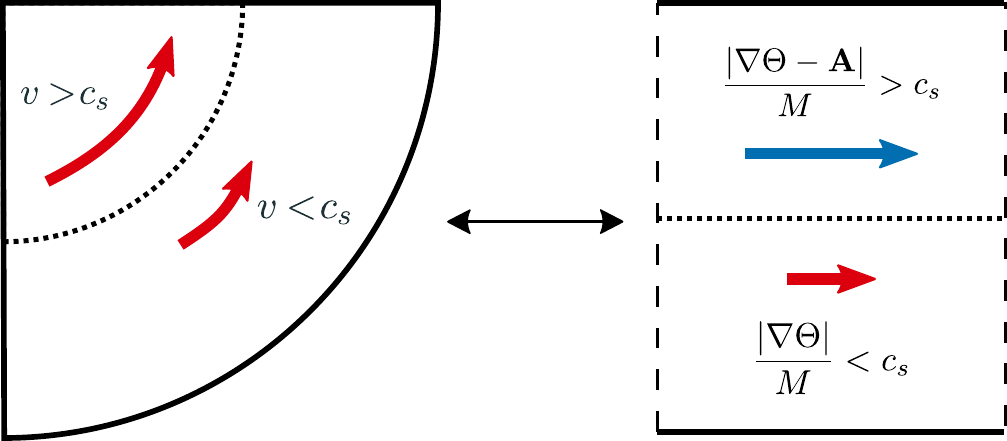}
  \caption{Schematic representation of out setup for the realization of an analogue ergosurface in a Cartesian geometry using synthetic vector potential fields.}
  \label{fig:setup}
\end{figure}

Let us start from the simplest case of a single planar ergosurface separating regions of sub- and super-sonic flow with a velocity directed along the $x$ axis parallel to the interface. As it is sketched in Figure \ref{fig:setup}, the synthetic vector potential gives the possibility to break the irrotationality constraint and generate a rotational superfluid flow using a two-dimensional BEC in the form of a plane wave of wavevector $\mathbf{K}$ with a spatially uniform density. The rotational flow is induced by a jump in the $x$ component of the synthetic vector potential, for instance from a vanishing value for $y<0$ to a finite value $A_x$ for $y>0$  (right panel).

The translational invariance of this geometry along $x$ offers crucial advantages from both the technical and the conceptual points of view. In particular, it guarantees that the $k_x$ component of the momentum is conserved during the scattering process. This is in analogy to the angular momentum conservation in cylindrical geometries (left panel). In contrast to the cylindrical case where the flow displays a non-trivial radial dependence, the asymptotic regions of our geometry are however flat and homogeneous, which allows to expand the incident and scattered waves in a plane wave basis. Finally, in contrast to the cylindrical geometry, the ergosurface is in our case an infinite line and the ergoregion an unbounded half plane. Thanks to this simpler geometry, the superradiance process at the ergosurface can be isolated from additional geometrical features and the products of the superradiant amplification are automatically evacuated without the need of a horizon.

It is interesting to note that geometrically similar flows can be studied in classical fluid mechanics. Also in this context, tangential discontinuities are known to show amplification for sound and gravity waves. An example is given in Section 84 of \cite{landau1987fluid} and is discussed in the context of superradiance \cite{brito2015superradiance}. But in the same book~\cite{landau1987fluid}, it is also pointed out that the amplification occurs together with the surface instabilities of shear flows, for example the Kelvin--Helmholtz instability.
As we are going to see in what follows, our system is instead immune to these instabilities, which leaves us with a stationary configuration in which superradiant scattering is the main physical process.

\subsection{Superradiant scattering in the hydrodynamic Klein-Gordon approximation}
\label{sec:superradiant_scatt}
As a first step, we consider the problem in the hydrodynamic limit and we derive a prediction for the amplification by means of a scattering approach~\cite{richartz2009generalized}. The impact of the superluminal features of the Bogoliubov dispersion will be discussed in Sec.\ref{sec:dispersion}.
In the configuration under examination, the acoustic metric has the form
\begin{equation}\label{eq:transverse-metric}
	g_{\m\n} = \frac{n_0}{c_s} 	
	\begin{bmatrix}
		-(c_s^2-v_x^2) & -v_x & 0\\
		-v_x & 1 & 0\\
		0 & 0 & 1
	\end{bmatrix},
\end{equation}
where the total physical velocity $v_x=v_x(y)$ includes the synthetic vector potential and is oriented along $x$. 

Because of the translational invariance along $x$ we can look for solutions of the form $\T_1(t,x,y)=e^{ik_x x}\p(t,y)$ with a well-defined $x$ component of the wavevector $\mathbf{k}$. Note that $\mathbf{k}$ is here the wavevector of the perturbation, to be distinguished from the one $\mathbf{K}$ of the underlying condensate.
Under this Ansatz, the Klein--Gordon equation \eqref{eq:kleingordon-cs} reduces the a single differential equation for $\p(t,y)$,
\begin{equation}\label{eq:kleingordon-transverseflow}
	-\left(\frac{1}{c_s}\de_t+i \frac{v_x}{c_s}k_x\right)^2\phi +\de_y^2\phi - k_x^2\phi=0.
\end{equation}
The analysis is further simplified if we restrict to cases where the flow velocity $v_x(y)$ has a $y$-dependence concentrated around $y=0$, while sufficiently far from this interface it acquires constant asymptotic values $v_x(y)=v_x^{s,f}$ in the \textit{slower} (s) region $y<0$ and in the \textit{faster} (f) one $y>0$. These velocities are related to the atomic canonical momentum $K_x$ and the synthetic vector potential by $v_x^s=\hbar K_x/M$ and $v_x^f=(\hbar K_x -A_x)/M$. For the sake of concreteness, we assume the velocities fulfill $0\leq v_x^s<v_x^f$, but our results are straightforwardly extended to other configurations.

\begin{figure}[t]
  \centering
  \includegraphics[width=\columnwidth]{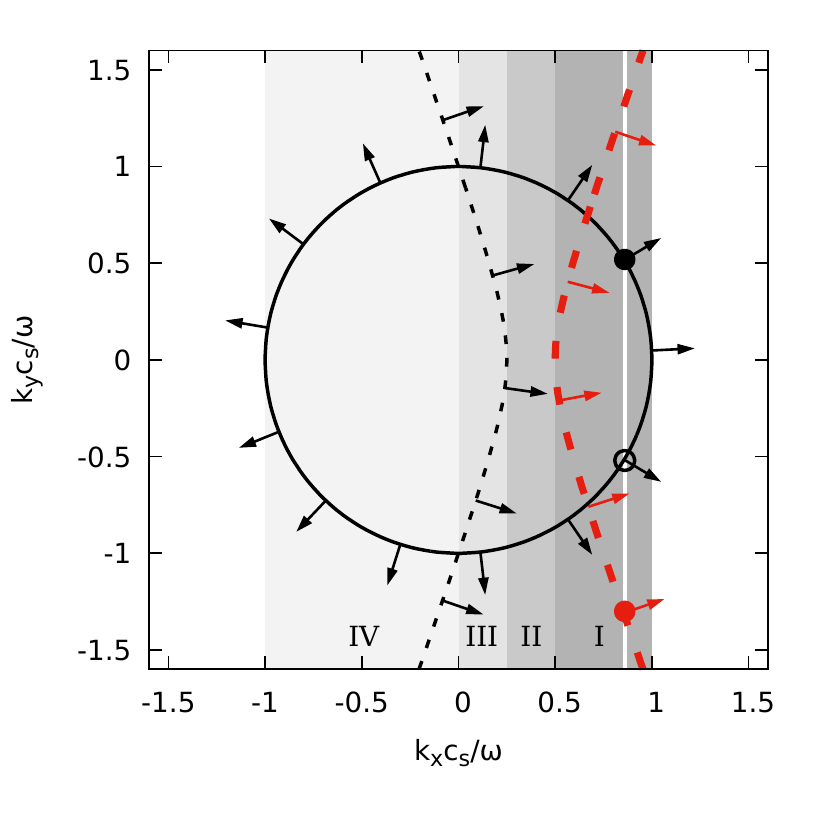}
  \caption{$\vc{k}$-space locus of modes at a given $\w$ for the Klein--Gordon equation in the \textit{slow} lower $y<0$ region (solid line) and the \textit{fast} upper $y>0$ region (dashed lines).  The slow region is taken at rest $v_x(y<0)=v_x^s=0$ and the fast one is moving with supersonic speed $v_x(y>0)=v_x^f=3c_s$.  The speed of sound $c_s$ is the same on both sides. Positive norm (see Appendix) branches are shown as thin black lines, negative norm branches are shown as thick red lines. For each mode the arrows indicate the direction of the group velocity $\vc{v}_g=\del_\vc{k}\w$. The different levels of gray (I-IV) indicate the $k_x$ intervals for which the different scattering processes occur (see text). The vertical white line in the darkest region (I) indicates an amplified superradiant scattering process, with the filled black dot indicating the incident mode, the empty black dot the reflected one and the red dot the transmitted one.
  }
  \label{fig:kg-dispersion}
\end{figure}

In particular, we look at the stationary scattering problem for an incident plane wave of frequency $\w$ coming from $y=-\infty$. In this case, we can expand the solution in plane waves as $\p(t,y)=e^{-i\w t}\vp(y)$ with $\vp(y\ll0)=e^{ik_y^s y} + \R e^{-ik_y^s y}$ and $\vp(y\gg 0)=\Tr e^{ik_y^f y}$. Within each region, the wavevector $k^{s,f}_y$ along $y$ is determined by the dispersion relations for the Klein--Gordon equation in the two regions
\begin{equation}
	\w = v^{s,f}_xk_x \pm c_s\sqrt{k_x^2+\left(k_y^{s,f}\right)^2},
	\label{eq:KG}
\end{equation}
where the plus and minus signs refer to positive- and negative-norm modes (see Appendix). It is immediate to analytically see that, for subsonic flows $c_s>v_x$, for a given positive frequency $\w>0$, only positive-norm modes are available and their $\vc{k}$-space locus has a closed, elliptic shape as shown by the solid lines in Fig.~\ref{fig:kg-dispersion}. For supersonic flows $c_s<v_x$, the locus consists instead of two hyperbolic branches of opposite norms (dashed lines in the same figure).

For given values of $\w$ and $k_x$, the $k_y$ values involved in the scattering process have to be chosen with the requirement that the group velocity $\vc{v}_g=\del_\vc{k}\w$ of the incident and transmitted waves has a positive $y$ component, so that the incident wave in the lower, slow region moves towards the interface and the transmitted wave in the upper, fast one moves away from it. For the same reason, the reflected wave in the lower region must be chosen in order for the group velocity to have a negative $y$ component. The fact that the flow velocity is parallel to the interface guarantees that the wavevectors of the incident and reflected waves have opposite $y$ components $\pm k_y^s$.

A concrete example of superradiant scattering process is given in Figure~\ref{fig:kg-dispersion} for a case where the lower $y<0$ region (solid lines) is subsonic and the upper $y>0$ one (dashed lines) is supersonic, so the $y=0$ interface is an ergosurface. The incident and reflected modes (filled and empty black dots) lie on a positive-norm (thin line) branch, while the transmitted mode (red dot) lies on a negative-norm (thick line) branch.

In order to establish the superradiant amplification, we can consider the relation between the reflection and transmission coefficients stemming from the conservation of the Wronskian $W = \vp (\vp^*)' - \vp' \vp^*$ between the solution and its complex conjugate. Physically, conservation of this Wronskian corresponds to the conservation of the current of norm of the modes along $y$. This provides a relation
\begin{equation}\label{eq:amplification-condition}
	1-\abs{\R}^2 = \frac{k_y^f}{k_y^s}\abs{\Tr}^2.
\end{equation}
between the reflection $\R$ and the transmission $\Tr$ amplitudes. From this relation it is immediate to see that the reflection coefficient exceeds one (i.e. the reflected wave is amplified) if the wavevectors $k_y^s$ and $k_y^f$ of the incident and transmitted waves have opposite signs. Given the form of our dispersion shown in Fig.\ref{fig:kg-dispersion}, this condition is naturally satisfied if the scattering process involves modes of opposite norms on the two sides. This is a sufficient condition for superradiant scattering to occur at an isolated ergosurface. A similar explanation for the amplification of waves at tangential discontinuities in classical fluid mechanics was given in \cite{mckenzie1972reflection}.

In addition to the superradiant amplified reflection discussed so far, other kinds of scattering processes can occur depending on the wavevector $k_x$, that is on the incidence angle from the lower region. The characterization of the different cases can be carried out by comparing the dispersion in the two regions as shown in Fig.~\ref{fig:kg-dispersion} and keeping in mind the conservation of $k_x$ at the interface~\cite{Russell1995}. The incident wavevector is to be chosen on the dispersion in the lower region (solid line). 

For instance, superradiant scattering is restricted to the darkest region (I) where a single, opposite norm mode is available for transmission in the upper region (thick red dashed line). In the neighboring, slightly lighter region (II), the incident wave is completely reflected since there is no available mode to transmit into the upper region. In the next two, even lighter regions (III-IV), ordinary scattering occurs and the incident wave is partially reflected and partially transmitted into a same-norm mode (thin black dashed line), the incident energy being distributed among the two in an incident-angle-dependent way as it happens for refraction of electromagnetic waves at the surface of a dielectric. While in all other regions (I-III) the $x$ component of the group velocity has the same positive sign in both the lower and upper regions, in region (IV) the incident and transmitted waves have opposite signs of the $x$ component of the group velocity, leading to a negative refraction phenomenon
~\cite{Veselago_1968}. In this case, the incident wave has a negative $x$ component of the group velocity, but due to the drag by the moving fluid,  the transmitted wave in the $y>0$ region deviates its path towards the positive-$x$ direction.

Whereas all other scattering process (II-IV) only involve positive norm modes and can also occur with non-uniform, yet everywhere subsonic velocity profiles, the superradiant process (I) crucially relies on the presence of a negative norm transmitted mode, which is only possible for a supersonic flow. To this purpose, it is worth noting that one cannot replace the change in the transverse velocity with a change in the local speed of sound, e.g. via 
a spatial modulation of the interaction constant as proposed in~\cite{balbinot2008nonlocal,carusotto2008numerical} for the analogue Hawking radiation. Even though negative norm modes emerge in the upper, fast region, superradiant scattering can not occur since there are no $k_x$ values for which states are simultaneously available on the positive-norm curve of the lower, slow region and on the negative-norm curve of the upper, fast region. This can be easily checked analytically. On a dispersion diagram such as Fig.\ref{fig:kg-dispersion}, it corresponds to the red thick dashed curve being always located to the right of the thin solid line.

As a final point, note how, in contrast to the cylindrical geometry, our translationally invariant (and thus Galilean invariant) geometry along $x$ gives symmetric roles to the upper and lower regions. As a result, amplification does not depend on the way in which the interface is crossed. In particular, the same superradiant scattering process occurs for wavepackets hitting the interface from the upper, fast region.

\subsection{Mapping to a 1D electrostatic problem}\label{sec:electrostatic}
The acoustic spacetime emerging from our translationally-invariant, two dimensional setup offers a realization of the \textit{rectilinear} model of the Kerr metric introduced in \cite{fulling1989aspects} as a toy model for the study of bosonic fields in rotating spacetimes. As it was explained there, in this case the problem of a massless neutral scalar field in the curved space-time can be reduced to a electrostatic problem in reduced dimensions. In this section, we take inspiration from these results to build an explicit link between our synthetic field configuration and the massive bosonic Klein paradox. This suggests a further direction in which our atomic system can be used as a quantum simulator of a relativistic quantum field theory.

\begin{figure}[t]
  \centering
  \includegraphics[width=\columnwidth]{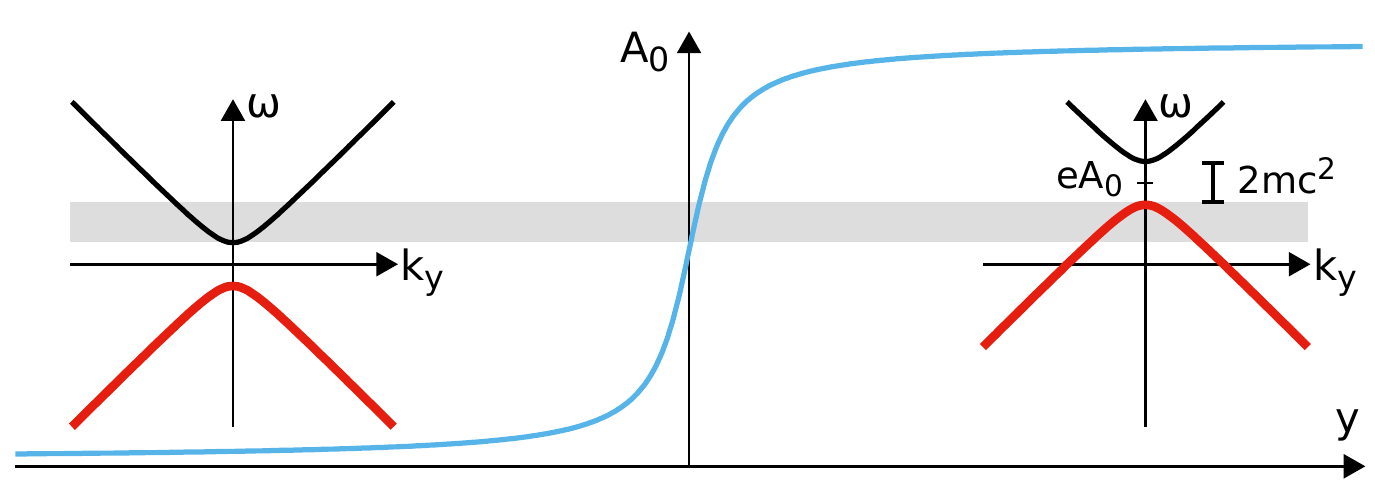}
  \caption{Schematic representation of the physical mechanism underlying the massive bosonic Klein paradox. The dispersion relations in the asymptotic regions are plotted along with the spatial profile of the electrostatic potential. The condition for the amplification to occur is to have both particle and antiparticle modes available at the same frequency, as indicated by the light gray shading in the plot.}
  \label{fig:kleinparadox}
\end{figure}

This link is easily understood by comparing the reduced one-dimensional Klein-Gordon equation for our analogue model \eqref{eq:kleingordon-transverseflow}  with the equation for a one-dimensional massive charged scalar field in an electrostatic potential $A_0$,
\begin{equation}
	-\left(\frac{1}{c}\de_t+i\frac{e}{\h c}A_0\right)^2\phi +\de_y^2\phi - \frac{m^2c^2}{\h^2}\phi=0.
\end{equation}
Here, $e$ is the charge, $m$ is the mass of the field and $c$ is the speed of light. The two equations are mapped into each other with the identifications
\begin{equation}\label{eq:parallelism}
	\frac{m^2c^2}{\h^2}\longleftrightarrow k_x^2 \hspace{0.8cm} \frac{eA_0}{\h}\longleftrightarrow k_x v_x \hspace{0.8cm} c\longleftrightarrow c_s\,:
\end{equation}
the role of the scalar potential $A_0(y)$ is played by the transverse velocity $v_x(y)$ and both the mass $m$ and the charge $e$ are controlled by the value of the transverse momentum $k_x$.
This parallelism makes it evident that superradiance in our system is equivalent to the bosonic Klein paradox, that is the amplified reflection of a bosonic field off an electrostatic potential step. In this analogy, the negative norm modes correspond to antiparticles, amplified reflection is associated to the transmission of antiparticles, and energy conservation corresponds to overall charge conservation. Because of the \textit{particle-hole symmetry} of the Bogoliubov problem, the positive frequency, negative-norm modes are actually physically equivalent to positive-norm modes at negative frequencies for $-k_x$. This is consistent with our identification of the transverse momentum $k_x$ with the charge of the particle in the reduced problem.

The condition for the superradiant process to happen can be derived by looking at the dispersion relations in the two asymptotic regions far from the potential step as shown in Figure \ref{fig:kleinparadox}. 
These plots correspond to a different \textit{cut} of the same dispersion relation studied in Figure \ref{fig:kg-dispersion}: there, the $\vc{k}$-space locus of modes at a given $\w$ was shown. Here, we plot instead the dependence of $\w$ on $k_y$ for a given $k_x$. It is immediate to see that the effect of a constant electrostatic potential is to rigidly shift the dispersion relation along the $\w$ direction. 

As a simplest example one can take the electrostatic potential $A_0(y)$ to be zero far in the $y< 0$ region and to tend to a positive constant $A_0$ far in the $y> 0$ region (thin blue line). If this value is large enough to satisfy
\begin{equation}\label{eq:condition-klein}
	eA_0>2mc^2\,,
\end{equation}
transmission to antiparticles, and hence amplification of the reflected wave, can occur in the range of frequencies $mc^2<\w<eA_0-mc^2$. The factor $2$ in the condition \eqref{eq:condition-klein} physically corresponds to the fact that a  particle-antiparticle pair is generated during the scattering process and both the particle and the anti-particle have the same mass $m$.

Through the identifications \eqref{eq:parallelism}, we can easily obtain from \eqref{eq:condition-klein} a necessary condition for amplified scattering,
\begin{equation}\label{eq:velocity-condition}
	\Delta v_x>2c_s
\end{equation}
where $\Delta v_x$ is the velocity jump across the interface, $\Delta v_x=v_x^f-v_x^s$. Quite remarkably, this condition shows that the presence of an analogue ergosurface separating sub- and super-sonic flows is not sufficient for superradiant scattering to occur, but a large enough velocity jump must be present. This condition is easily understood based on the Galilean invariance of our setup under velocity boosts along the $x$ direction: as long as $v_x<2c_s$, there exists in fact a reference frame in which the flow is everywhere subsonic and superradiance can not occur.

While the parallelism with the Klein paradox has been rigorously established for a given $k_x$, it is important to keep in mind that the non-trivial form of the identifications in \eqref{eq:parallelism} make that our sound scattering process to have a completely different angular $k_x$-dependence  from the one of a charged field hitting a scalar potential step at different angles. 
For instance, for $k_x=0$ waves at normal incidence, the mass gap in the corresponding electrostatic problem vanishes. Since for a massless field there is no forbidden mass gap to overcome and every non-zero electrostatic field provides amplification within a suitable interval of incident frequencies, one could have expected amplification to occur for every small value of $v_x$. However, one must also remember that for $k_x=0$ also the charge vanishes in our identification, so the scalar potential has no effect and superradiant scattering is forbidden.

Finally, it is worth noting that condition \eqref{eq:velocity-condition} is the same result derived in \cite{landau1987fluid} for the hydrodynamic tangential discontinuity. Notice however that, because of the rotational velocity field, the metric treatment of that system (such as the one presented in \cite{brito2015superradiance}) is not a full description of the physics because of the presence of surface unstable modes.

\subsection{The role of the  superluminal Bogoliubov dispersion}\label{sec:dispersion}
Up to now we have considered the problem within the hydrodynamic approximation, where sound is described by a Klein--Gordon equation and hence displays a linear dispersion relation. In reality, the collective excitations in a uniform flowing BEC follow a Doppler-shifted version of the celebrated Bogoliubov dispersion~\cite{pitaevskii2016bose}, 
\begin{equation}\label{eq:bogo-dispersion}
	\h\w=\h \mathbf{v}\cdot \mathbf{k} \pm\sqrt{\frac{\h^2k^2}{2M}\left(\frac{\h^2k^2}{2M}+2gn\right)}.
\end{equation}
This dispersion is well linear and sonic at small momenta but then grows quadratically at higher momenta, i.e. becomes \textit{superluminal}. The first term accounts for the Doppler shift of the frequency when moving from the fluid to the laboratory frame.

While the superluminal nature of the Bogoliubov dispersion is not expected to completely suppress the superradiance effects, important modifications may well appear. The first study of superradiant scattering for dispersive fields~ \cite{richartz2013super} focused on the Klein paradox for a one-dimensional massless field. In what follows, we extend the study to the general two-dimensional case of a Bogoliubov dispersion. In the previous subsection we have seen how the transverse momentum  $k_x$ of the sound waves, besides giving the coupling with the background flow, also provides a mass term for the dimensionally reduced Klein paradox problem. We are now going to show how the main effect of the superluminal dispersion will be encoded in a modification to this mass term.

\begin{figure}[t]
  \centering
  \includegraphics[width=\columnwidth]{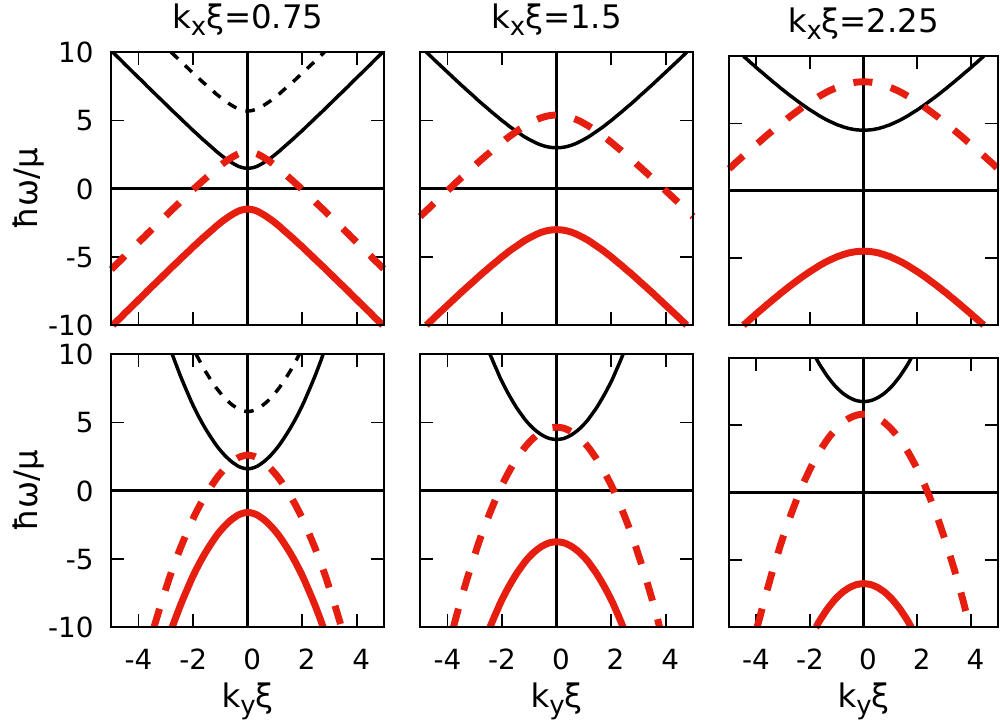}
  \caption{Constant-$k_x$ cuts of the hydrodynamic Klein--Gordon (upper panels) and the Bogoliubov (lower panels) dispersion relations  for increasing values of $k_x$ (left to right). Solid lines refer to the slow region at rest with $v_x^s=0$ and dashed lines to the fast region with $v_x^f=2.8 c_s$. Black (thinner) lines are the positive-norm branches, red (thicker) lines are the negative-norm ones. In the Klein--Gordon case, if the condition \eqref{eq:velocity-condition} is satisfied, superradiant scattering remains possible at all $k_x$ thanks to the enduring intersection between the positive-norm branch in the slow region and the negative-norm one in the fast region (upper panels). In the Bogoliubov case, this intersection is only present for low or moderate values of $k_x$ (bottom-left and bottom-center panels) and disappears for high enough values of $k_x$ for which superradiant scattering is no longer possible (bottom-right panel).}
  \label{fig:dispersion-suppression}
\end{figure}

Within the one-dimensional perspective at a given $k_x$, having a finite frequency range for superradiance requires that the maximum of the negative norm branch in one region be higher than the minimum of the positive norm branch in the other region. Imposing this requirement on the Doppler-shifted dispersion relations \eqref{eq:bogo-dispersion} for velocities parallel to $x$ implies that
\begin{equation}
	\h\, \Delta v_x\, k_x-2\sqrt{\frac{\h^2k_x^2}{2M}\left(\frac{\h^2k_x^2}{2M}+2gn\right)}>0,
\end{equation}
where we have taken $\Delta v_x=v_x^f-v_x^s>0$. This condition is satisfied for
\begin{equation}
	0<\frac{\h^2 k_x^2}{M^2}<\Delta v_x^2-4c_s^2,
\end{equation}
which implies that, in contrast to the Klein--Gordon case, there is a maximum value of the transverse momentum above which superradiance can no longer occur,
\begin{equation}
k_x^{\rm max} = \frac{\sqrt{\Delta v_x^2 - 4c_s^2}}{c_s}\,\xi^{-1}.
\end{equation}
The dependence of this upper bound on $\xi$ highlights the dispersive origin of the effect. As it is illustrated in Figure \ref{fig:dispersion-suppression}, this upper bound can be related to the nonlinear $k_x$-dependence of the effective mass gap in the reduced one-dimensional Bogoliubov problem. For small $k_x$, both the mass gap and the rigid upward shift of the dispersions given by the effective electrostatic potential grow linearly in $k_x$. For large $k_x$, the mass gap grows faster than the rigid shift, so the negative- and positive-norm modes eventually stop intersecting for large enough values of $k_x$.

\begin{figure}[t]
  \centering
  \includegraphics[width=\columnwidth]{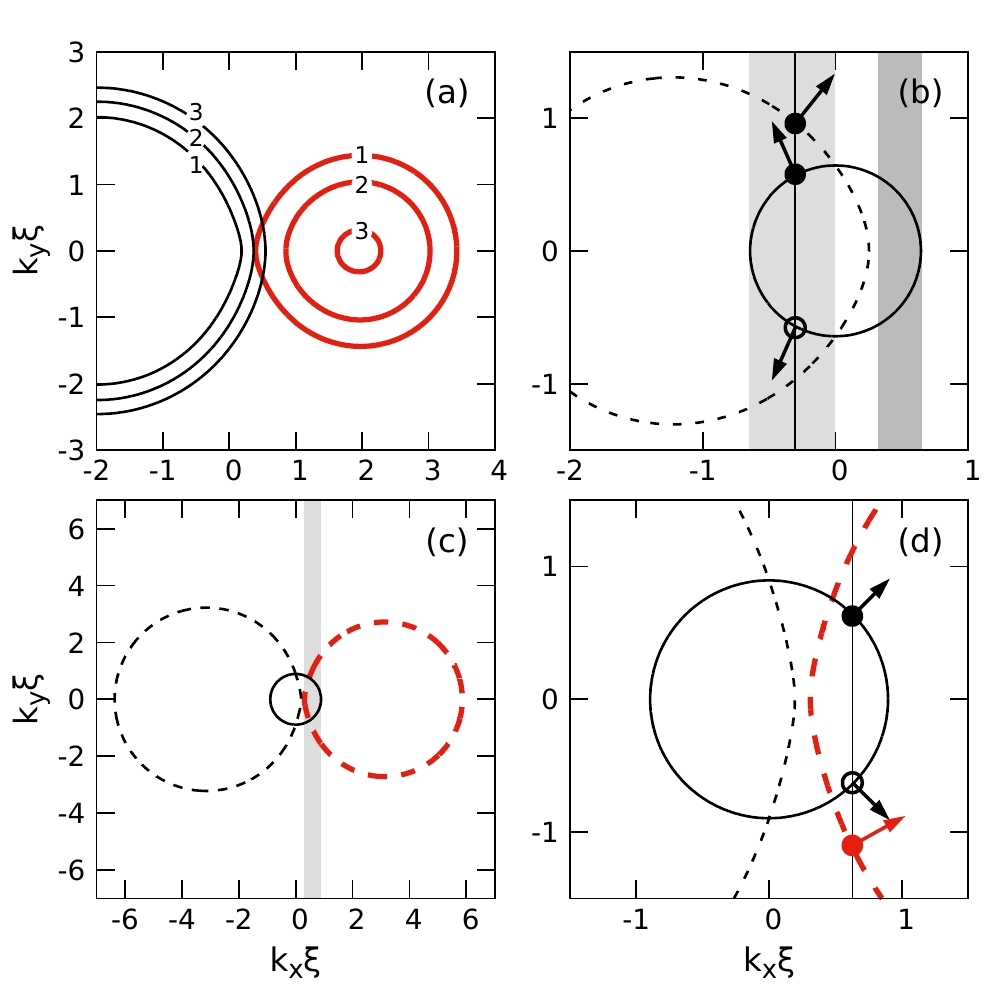}

  \caption{Cuts of the Bogoliubov dispersion relation at constant $\w$. Solid (dashed) lines refer to the slow (fast) region at $y<0$ ($y>0$). The speed of sound $c_s$ is the same on both sides. Black (thin) lines are positive-norm modes and red (thicker) ones negative-norm ones. Arrows indicate the direction of the group velocity, that is outward for the black curves and inward for the red ones.
  (a) Cuts at different values of $\h\w/Mc_s^2$ (indicated by the numbers on the curves) in a uniform region with a supersonic speed $v_x^f=2c_s$: because of the superluminal dispersion these curves have an oval shape also in the supersonic case, rather than the hyperbolic one of the Klein--Gordon case.
  (b) Dispersions at $\h\w=Mc_s^2$ in the two zones for a slow region at rest $v_x^s=0$ and a fast region with a subsonic speed $v_x^f=0.83\,c_s$. In the darker gray region one has total reflection while in the lighter gray one \textit{negative refraction} occurs; the dots and the arrows indicate the modes involved in an example of such process and their group velocities. This process will be addressed in the GPE simulation shown in the lower panels of Fig.\ref{fig:gpe-evolution-packet}.
  (c) Dispersions at $\h\w=1.5Mc_s^2$ for a slow region at rest $v_x^s=0$ and a fast one with a supersonic speed $v_x^f=\pi c_s$: in the gray region superradiant scattering is possible. (d) A closeup of the same plot; the dots and the arrows indicate the modes involved in an example of such process and their group velocities. This specific process will be addressed in the GPE simulations shown in Fig.\ref{fig:gpe-evolution} and in the upper panels of Fig.\ref{fig:gpe-evolution-packet}.}
  \label{fig:bogo-cases}
\end{figure}

Further light on this physics can be obtained by looking at the constant-$\w$ cuts of the Bogoliubov dispersion relation \eqref{eq:bogo-dispersion} that are displayed in the different panels of Figure \ref{fig:bogo-cases}.
Panel (a) shows how the main effect of the superluminal dispersion is to change the shape of the curves in the supersonic region: instead of the hyperbolic open shape of the Klein--Gordon case shown as dashed lines in Figure \ref{fig:kg-dispersion}, they now have a closed, oval-like shape. For increasing $\w$, the oval corresponding to the positive norm modes expands, while the one corresponding to the negative norm ones shrinks and eventually disappears above some critical frequency.

Analogously to the Klein--Gordon case presented in Figure \ref{fig:kg-dispersion}, the occurrence of superradiant scattering can be visualized from the intersection of both positive- and negative-norm curves with the vertical line at a fixed $k_x$: this happens in the gray region in panel (c) and an example of such process is displayed on a magnified scale in panel (d). As before, also other kinds of scattering behaviours can be recognized depending on parameters: in the darker gray region of panel (b) the incident wave gets totally reflected, while in the lighter gray region negative refraction occurs. 

While these cases exhaust the possibilities for a velocity parallel to $x$, in the following Sec.\ref{sec:horizon} we shall see how even more complex scattering processes can occur once the flow can acquire also a $y$ component.

\subsection{GPE numerical calculations}\label{sec:gpe}

In order to shine further light on superradiant scattering and confirm the predictions drawn from the graphical study of the analytical dispersion relations, we performed numerical simulations of the time dependent dynamics of the two-dimensional GPE \eqref{eq:gpe_gauge}. For the background condensate, we take a real and constant order parameter $\Psi_0$ and a spatially uniform interaction strength, so that the canonical velocity is zero and the speed of sound is spatially uniform and equal to $c_s=\sqrt{gn/M}$. We take the vector potential directed along $x$ with $A_x(y<0)=0$ and $A_x(y\geq0)=A_x$ constant, so to give a sudden jump in the physical velocity along $x$. In order to maintain the plane wave shape of the BEC at all times, we need to introduce an external potential jump $V(y\geq0)-V(y<0)=-A_x^2/(2M)$. We impose periodic boundary conditions in both directions and we ensure that the integration box is large enough for finite size effects to be irrelevant for the computational times of interest~\cite{giacomelli2019ergoregion}.

Among many other possible schemes that may be implemented, our choice of using a vector potential that is directed along $x$ and only varies along $y$ is beneficial from both the experimental and numerical point of view. Such a configuration could be, for instance, obtained by means of a pair of counterpropagating Raman laser beams directed along the $\pm x$ directions and a $y$-dependent magnetic field that varies the detuning of the atomic states~\cite{spielman2009raman}.

Time evolution is numerically performed with a split-step pseudo-spectral method, in which we apply the propagator of the GPE as $e^{-iHt}\sim e^{-i\tld{V}t/2}e^{2iA_x(y)\,p_x}e^{-ip^2/2M}e^{-i\tld{V}t/2}$, where $\tld{V}$ contains all the terms multiplicative in position space such as the external potential. As usual, the kinetic energy is included in a multiplicative way in Fourier space for both $x,y$. Thanks to the specific form $A_x(y)$ of the vector potential considered here, also this latter can be included in the calculation as a multiplicative factor provided the Fourier transforms along $x$ and $y$ are performed separately and the vector potential is applied in between to the partially Fourier transformed wavefunction $\tilde{\Psi}(k_x,y)$.

In order to study superradiance phenomena, we impose on top of the uniform condensate a small amplitude ($|\delta \Psi/\Psi_0|\ll 1$) wavepacket with a plane wave shape of wavevector $k_x$ along $x$ and a Gaussian shape along $y$ with a carrier wavevector $k_y$. The variance $\sigma_y$ is taken sufficiently small for the wavepacket to be localized in the $y<0$ region, but large enough for the momentum distribution to be sharply peaked around the desired $k_y$.   The central wavevector $(k_x,k_y)$ is chosen to be on the positive-norm mode of the slower region with a group velocity directed towards the interface. In order to obtain a clean wavepacket of Bogoliubov excitations, positive and negative frequency components in the atomic basis must be suitably combined to only have positive frequencies in the Bogoliubov quasiparticle basis.

The value of the vector potential is chosen so to satisfy the condition \eqref{eq:velocity-condition} for amplification.
In particular, the same parameters of Figure \ref{fig:bogo-cases}(d) are used; the chosen wavevector $(k_x,k_y)$ is indicated there as a black dot.
Since $k_x$ and $\w$ are conserved, we expect that the wavepacket is transmitted in the faster region on the mode indicated by a red dot and located on the negative-norm (red) oval. At the same time, the amplified reflected wavepacket is expected to appear at the wavevector indicated by the black empty dot. 

\begin{figure}[t]
  \centering
  \includegraphics[width=0.8\columnwidth]{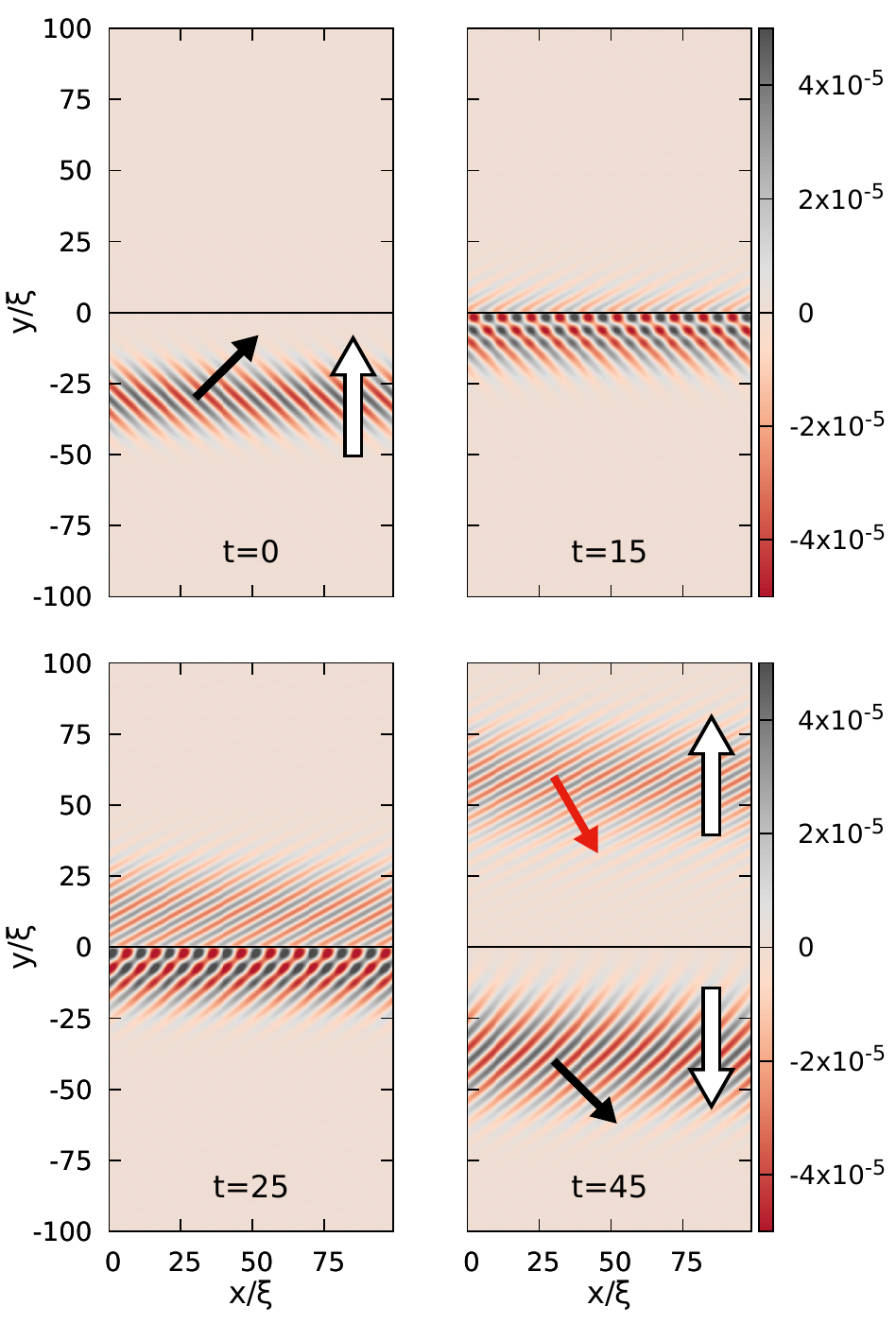}
  \caption{Snapshots of the 
 time-evolution of a superradiant scattering process as predicted by a numerical simulation of the GPE. The color plot shows the spatial profile of the density modulation with respect to the uniform unperturbed BEC. We impose a perturbation wavepacket on top of a uniform BEC in the presence of a vector potential $A_x=-\pi Mc_s$ in the $y>0$ half-plane and a compensating external potential as specified in the text. The interaction constant $g$ is spatially uniform, so to give a constant speed of sound $c_s$. The initial gaussian wavepacket has $k_x=0.63/\xi$, a central $k_y=0.63/\xi$ (indicated as dot in the dispersion plot of Fig.\ref{fig:bogo-cases}(d)) and is spatially centered around $y=-30\xi$ with a variance $\sigma_y=8\xi$. Times are expressed in units of $\mu/\h$. The white arrows indicate the group velocity of the wavepackets along $y$. The thick black (red) arrows indicate the directions of the wavevector (i.e. of the phase velocity) of the  positive (negative) norm wavepackets. One can recognize the negative norm wavepacket from the opposite directions of the group velocity and of the wavevector along $y$. The simulations have been carried out in an integration box of size $L_x=20\xi$ along $x$ and $L_y=200\xi$ along $y$. Grid spacings $\Delta x=\Delta y=0.2\xi$ are taken and a time step $\Delta t = 10^{-3}\mu/\h$. 
 Thanks to the periodic boundary conditions along $x$, visibility of the figure was improved by expanding the $x$ domain by repeating the data multiple times.}
  \label{fig:gpe-evolution}
\end{figure}

Snapshots of the time evolution for parameters for which one expects amplification are shown in Figure \ref{fig:gpe-evolution}. For each wavepacket, the white arrows indicate the direction of the group velocity along $y$ while the red and black arrows indicate the directions of the phase velocities. We can recognize the negative norm transmitted wavepacket in the upper region of the last snapshot from the fact that the $y$ components of the phase and group velocities have opposite signs, as expected from the dispersion diagram of Fig.\ref{fig:bogo-cases}(d). For all numerical wavepackets, a Fourier analysis confirms that their central wavevectors match the ones expected from the analytical dispersion relation shown in Fig.\ref{fig:bogo-cases}(d).

\begin{figure}[t]
  \centering
  \includegraphics[width=\columnwidth]{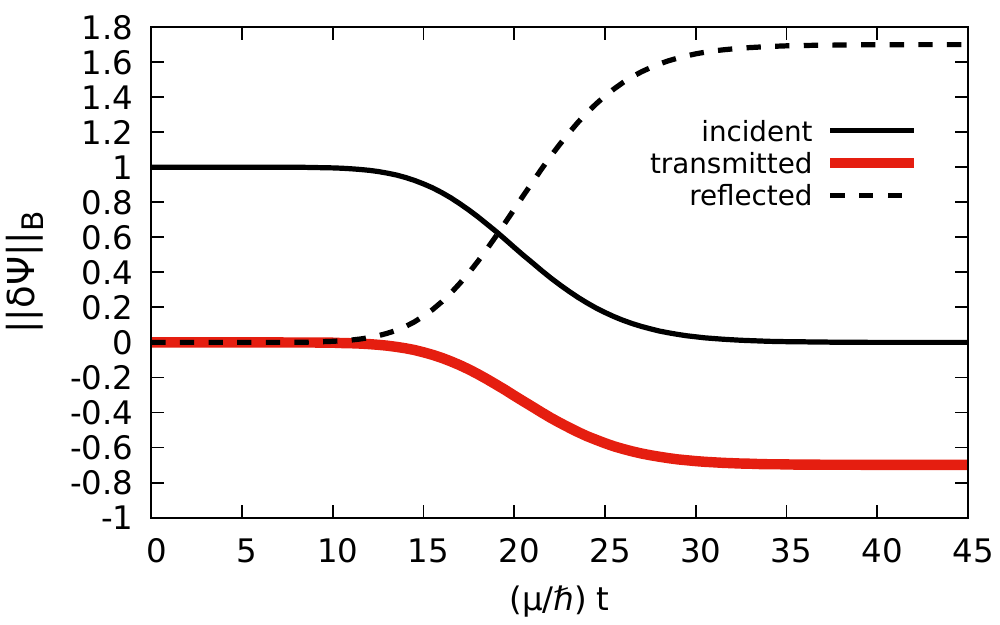}
  \caption{Time dependence of the Bogoliubov norm \eqref{eq:packet-bogo-norm} of the wavepackets, normalized to the one of the initial in-going wavepacket, for the parameters of the simulation of Figure \ref{fig:gpe-evolution}. One can see that the reflected wavepacket is amplified by roughly $70\%$.}
  \label{fig:integrated-packets}
\end{figure}

\begin{figure*}[t]
  \centering
  \includegraphics[width=\textwidth]{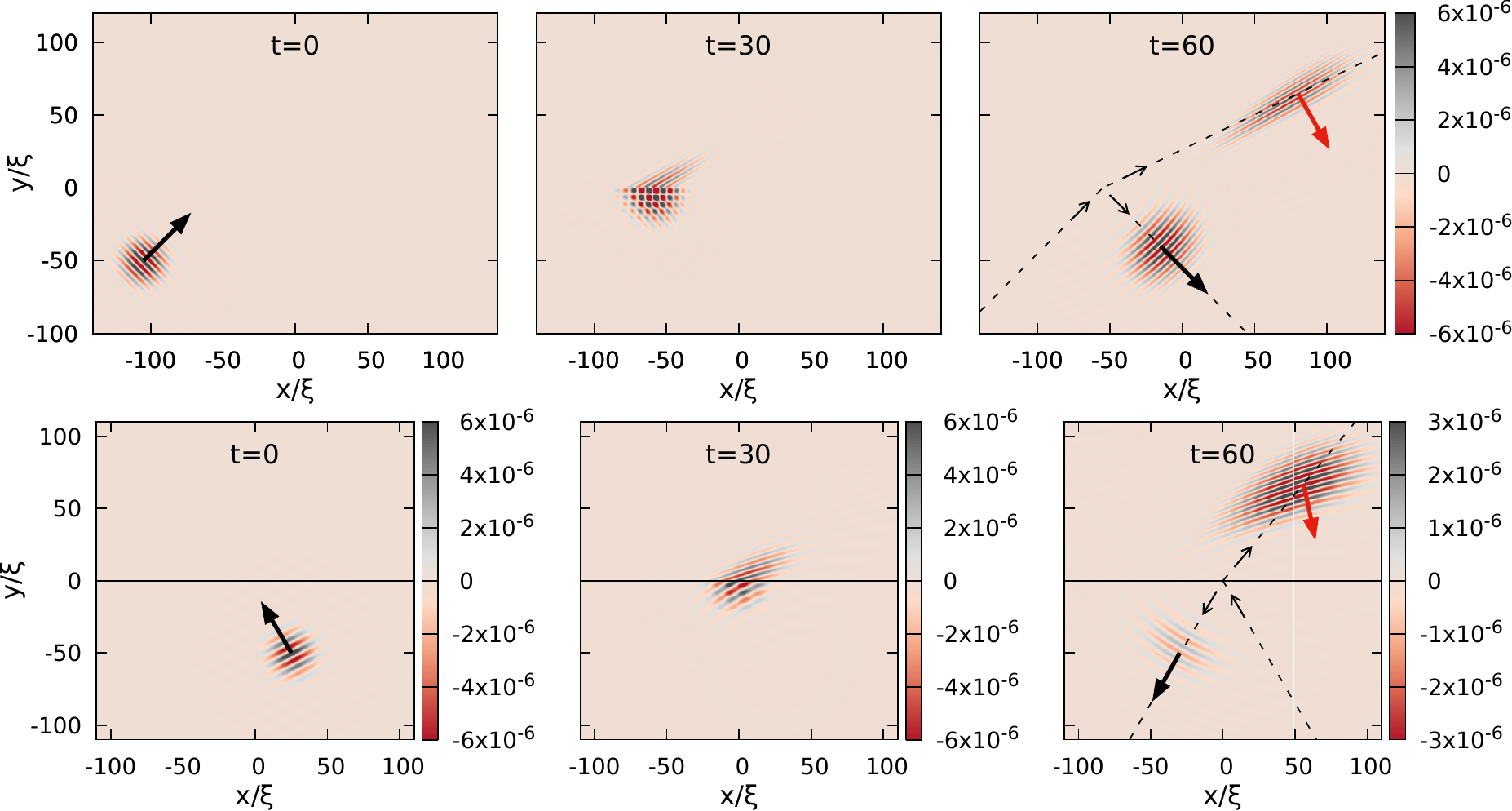}
  \caption{Upper panels: snapshots of the scattering process for the same configuration as in Figure \ref{fig:gpe-evolution} except for a finite width $\sigma_x=\sigma_y=10\xi$ of the wavepacket along $x$. The integration box has the same size $L_y=200\,\xi$ along $y$ but a wider size $L_x=200\,\xi$ along $x$. The color plot shows the spatial profile of the density modulation with respect to the uniform unperturbed BEC. The thick black and red arrows indicate the direction of the wavevector (i.e. of the phase velocity) of the positive-norm and negative-norm wavepackets, respectively. The dashed lines in the rightmost panels are the trajectories of the center of the wavepackets during the scattering and the thin arrows on them indicate the direction of the group velocities.
  Lower panels: analogous plots in the case of negative refraction. The vector potential in the upper region is $A_x=-1.26\;Mc_s$ and the initial gaussian wavepacket is centered in $k_x=0.3/\xi$ and $k_y=0.58/\xi$ and has variances $\sigma_x=\sigma_y=10\;\xi$. Notice the different scale used in the bottom-right panel, showing the reduced amplitudes of the reflected and negative-refracted wavepackets as compared to the superradiant case shown in the upper-right panel.}
  \label{fig:gpe-evolution-packet}
\end{figure*}

To numerically verify that the expected amplification of the reflected wavepacket is indeed taking place one cannot simply look at the maximum of the wavepackets, since the presence of superluminal dispersion leads to a spreading of the wavepacket. One can instead compute the Bogoliubov norm of the wavepackets (see the Appendix), that is conserved by the time evolution of linear perturbations and corresponds to a generalized number conservation in which negative-norm modes are weighted with a minus sign:
\begin{equation}\label{eq:packet-bogo-norm}
	||\delta\psi||_B(t) = \int \mathrm{d}k_y\left(\abs{u(t,k_y)}^2-\abs{v(t,k_y)}^2\right).
\end{equation}
Here $u$ and $v$ are the positive and negative frequency components of the wavepacket in the atomic basis and in the fluid reference frame. In practice, these two components can be isolated by computing the spatial Fourier transform of the two regions separately and identifying the components of wavevector $\pm k_x$. For our choice of a plane-wave along $x$, the positive and negative $k_x$ wavevector components are in fact directly associated to the positive and negative frequency ones of \eqref{eq:packet-bogo-norm}. Within each region, the in-going and out-going wavepackets can be isolated since the peak of their momentum distributions is located at values of $k_y$ with opposite signs: for example the in-going initial wavepacket will have its positive frequency peak at $(k_x,k_y^s)$ and its negative frequency one at $(-k_x,-k_y^s)$, while the out-going reflected one will have them respectively at $(k_x,-k_y^s)$ and $(-k_x,k_y^s)$. 

The resulting time dependence of the norms of the different wavepackets is shown in Figure \ref{fig:integrated-packets}, where one can see that the reflected wavepacket is strongly amplified by approximately  $70\%$ as compared to the incident wavepacket. The negative value of the norm of the transmitted wavepacket exactly corresponds to the amplification, so that the total norm and energy are correctly conserved in the scattering process. This confirms the physical interpretation that the amplification of the reflected wavepacket is compensated by the storage of negative energy in the upper region.

While using a plane wave form along $x$ was beneficial to draw the analogy with the electrostatic problem and perform a quantitative study of the time-evolution of the norm, a clear intuitive picture of the scattering process can be obtained by performing an analogous calculation for a wavepacket of finite width also along $x$. Even though the mapping onto the electrostatic problem is no longer exact, the same energetic considerations hold. The result is shown in the upper panels of Figure \ref{fig:gpe-evolution-packet}, where the thick arrows again point in the direction of the phase velocity of the wavepacket while the thin ones along the dashed lines of the last panel indicate the direction of the group velocity. Even though the overall geometry of the scattering process closely resembles standard refraction, a computation the norm of the different wavepackets shows that the reflected wavepacket has indeed been amplified. 

It is interesting to compare this scattering process with the negative refraction process~\cite{Veselago_1968} that takes place for different values of the incident wavevector and of the vector potential (lower panels). In this case, depicted in Fig.\ref{fig:bogo-cases}(b), the incident wavepacket has a negative $x$-component of the group velocity, but the transmitted wavepacket is dragged back by the moving condensate towards the positive-$x$ direction. Since no amplification is taking place, the reflected and transmitted intensities sum up to the incident one and the reflected and transmitted wavepackets are individually weaker than the incident one.

As a final remark, we need to emphasize that these simulations indicate that we are dealing with superradiant scattering from a dynamically stable interface: it is clear from the numerical GPE simulations that the interface quickly returns to its unperturbed state once the wavepackets have moved away from it.
This numerical result will be confirmed by the Bogoliubov analysis presented in Sec.\ref{sec:modematch}, where we find no unstable modes on the surface. This is a key difference from the case of tangential discontinuities in hydrodynamics discussed in~\cite{landau1987fluid}. In this case, discontinuity surface is typically dynamically unstable and tends to quickly develop ripples that complicate the observation of superradiant amplification.

\section{Superradiant dynamical instabilities}\label{sec:instabilities}

We have seen that superradiant scattering processes discussed in the previous Sections crucially rely on the presence of negative energy modes in some part of the system; the presence of such modes is called \textit{energetic instability}. In our configuration, the negative energy modes are available in the upper half of the system and can be populated while conserving the energy by radiating positive energy away in the lower half. Still, in the case of a single interface considered so far, this process cannot happen spontaneously (at least at the classical level) and it must be continuously stimulated by some incident wave.

Things change if different boundary conditions are considered. Take for example a configuration in which, instead of having an unbounded system which can evacuate waves in both the $\pm y$ directions, we introduce a reflecting boundary condition for fluctuations at the upper system edge. In this case, the transmitted negative norm wavepacket will get reflected and sent back to the interface. Since amplification does not depend on the way the interface is crossed, amplified superradiant reflection will now take place in the upper part, a stronger wavepacket will be generated that propagate upwards, and the process will continue indefinitely. This bouncing back and forth of the wavepacket between the interface and the reflecting boundary is associated to sizable amplification at each bounce on the interface, so that the amplitude of the \textit{trapped} negative-energy mode will increase indefinitely until saturation effects beyond our Bogoliubov model start taking place. 

\begin{figure*}[t]
  \centering
  \includegraphics[width=\textwidth]{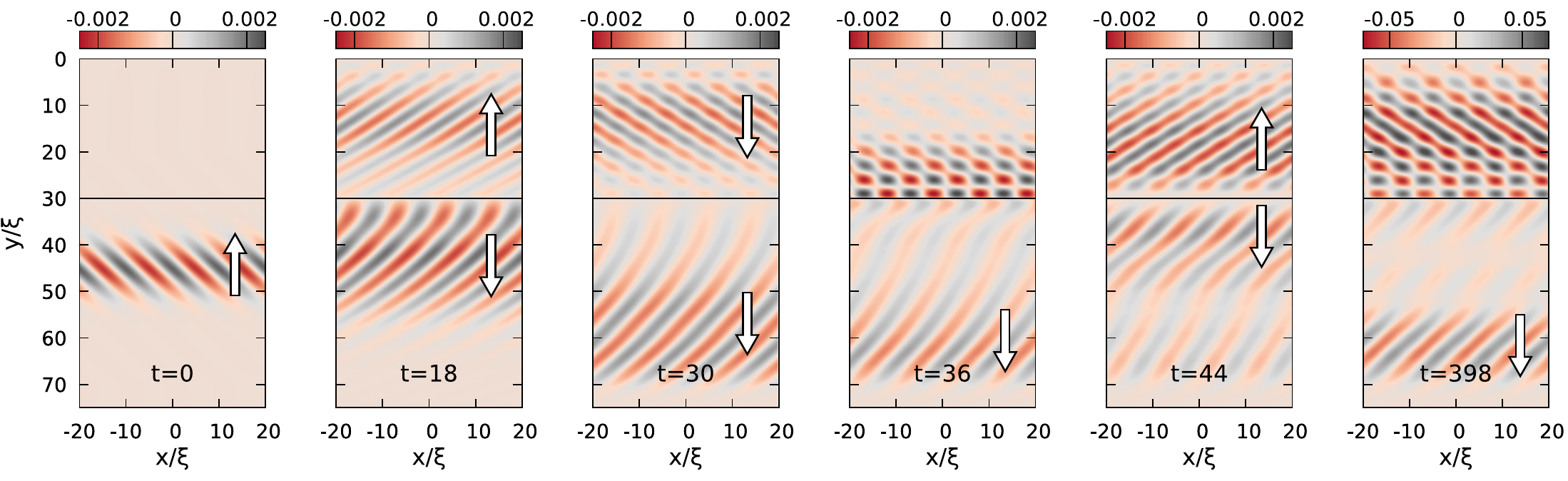}\\
  \includegraphics[width=\textwidth]{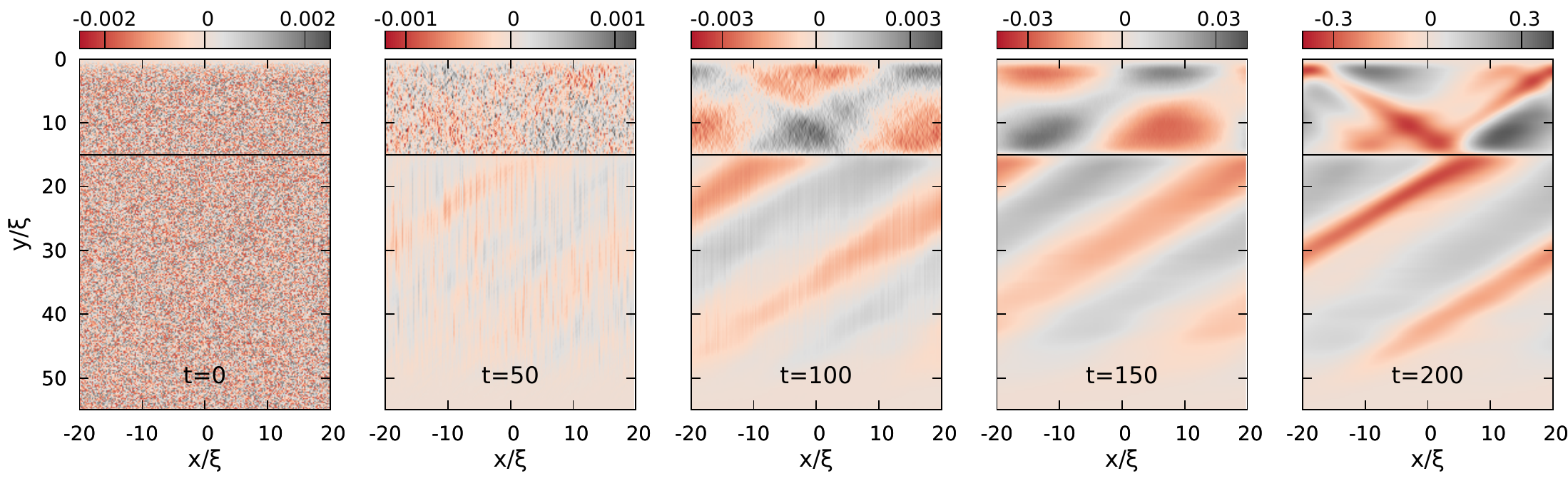}
  \caption{Snapshots of the time evolution of a dynamically unstable condensate as predicted by a numerical GPE calculation starting from different initial states. The condensate is confined along $y$ in a box of length $y_{max}=80\xi$. The interaction constant $g$ is constant, giving a spatially uniform speed of sound $c_s$. In the $0\le y\le L$ region, a transverse vector potential $A_x=-3Mc_s$ and a compensating external potential $V=-A_x^2/(2M)$ are applied. Fluctuations are absorbed via an imaginary potential when approaching the lower edge at $y_{max}$, so to mimic an open system geometry along that direction. On the upper row, the initial state features an incident wavepacket traveling in the upwards direction towards the \textit{cavity} of length $L=30\xi$; the white arrows indicate the directions along $y$ of the group velocities of the wavepackets. On the lower row, the initial state only features a weak white noise and the cavity length is $L=15\xi$.}
  \label{fig:gpe-evolution-noise}
\end{figure*}

In the general relativity analogy, this configuration can be seen as an analog of the ergoregion instability of a fast-spinning star with no horizon~\cite{brito2015superradiance}. The spacetime region within the ergosurface shows an exponentially growing perturbation, while correspondingly growing waves get emitted into the outer space. But the {\em dynamical instability} mechanism is not restricted to this specific geometry: an analogous dynamics would, e.g., occur if the reflecting boundary condition was imposed in the lower part of the system, resulting in an exponential growth of a trapped positive-energy mode. In the general relativity context, this situation can be associated to the black hole bomb instabilities that occur when amplified (positive-energy) waves are sent back to the ergosurface by some effective external \textit{mirror}~\cite{brito2015superradiance}, while the negative-energy waves falls through the horizon into the black hole. Qualitatively similar instabilities take place also in more complex configurations with \textit{stripe}-shaped regions of fast motion within a medium at rest.

In the  electrostatic model, these configurations correspond to a square box electrostatic potential for the Klein--Gordon equation. This is known to give rise to dynamical instabilities as first derived in \cite{schiff1940existence} with a toy model for nuclear physics in mind, and has been thoroughly investigated in \cite{fulling1989aspects,lv2016bosonic}, also in connection with rotating spacetimes. The emergence of these dynamical instabilities is known in the literature as the Schiff--Snyder--Weinberg (SSW) effect.

\subsection{Numerical simulations}
\label{sec:instab_nume}

As a concrete example of this physics, we consider a finite condensate trapped along the $y$ direction in a box potential; the vanishing density at the upper boundary at $y=0$ introduces reflecting boundary conditions for the Bogoliubov excitations. An absorbing region for fluctuations is introduced around the lower boundary at $y_{max}$ so to simulate an open system geometry in this direction.
We then apply a transverse synthetic vector potential field in the upper region $y\in[0,L]$, with $L<y_{max}$. 

In the upper panels of Figure \ref{fig:gpe-evolution-noise}, we display the numerical solution of the Gross--Pitaevskii equation for this configuration starting from an incident wavepacket traveling in the upwards direction with a wavevector in the superradiant amplification range. At the first bounce on the $y=L$ interface, an amplified reflected wavepacket is obtained via superradiant scattering. The negative-norm transmitted wavepacket keeps bouncing back and forth between the interface at $y=L$ and the reflecting boundary at $y=0$ while its intensity exponentially grows.

In the lower panels of Figure \ref{fig:gpe-evolution-noise} we display an analogous numerical simulation starting from a noisy initial state. In this case, the development of the dynamical instability appears qualitatively different. In the $y>L$ lower region, one can see the emergence of a pattern that can be recognized as a \textit{down-going} wave, while a stationary wave coming from superposition of \textit{up-going} and \textit{down-going} waves appears in the upper region with an exponentially growing amplitude. This latter standing wave is the trapped negative-energy mode that gets self-amplified while a positive-energy wave is radiated in the downwards direction.

\subsection{The Bogoliubov spectrum}
\label{sec:instab_Bogo}

Further light on these phenomena is offered by a study of the Bogoliubov eigenmodes. This is done by linearizing the GPE around some stationary state $\Psi_0$ as $\Psi=\Psi_0+\delta\Psi$. Thanks to the translational symmetry along $x$, we can adopt the one-dimensional perspective and study the spectrum of the excitations for fixed transverse momentum $k_x$. This corresponds to taking a plane wave form $\delta\Psi(x,y,t)=e^{ik_x x}\vp(y,t)$ for the fluctuations; the field $\vp$ then satisfies the one-dimensional Bogoliubov problem~\cite{castin2001bose}
\begin{equation}\label{eq:bogo-problem-gauge}
	i\h\de_t 
	\begin{pmatrix}
		\vp\\\vp^*
	\end{pmatrix}
	=
	\begin{bmatrix}
		D_+ & g\abs{\Psi_0}^2\\
		-g\abs{\Psi_0}^2 & -D_-
	\end{bmatrix}
	\begin{pmatrix}
		\vp\\\vp^*
	\end{pmatrix},
\end{equation}
where the field $\vp$ and its complex conjugate are treated as independent variables and
\begin{equation}
	D_\pm=-\frac{\h^2}{2M}\de_y^2+\frac{(\pm k_x-A_x)^2}{2M}+2g\abs{\Psi_0}^2+V-\mu\,.
\end{equation}
The spectrum of this problem can be studied by diagonalizing the equation \eqref{eq:bogo-problem-gauge} in matrix form. Since the matrix is not hermitian, complex eigenvalues can arise (see the Appendix for more details).

\begin{figure}[t]
  \centering
  \includegraphics[width=\columnwidth]{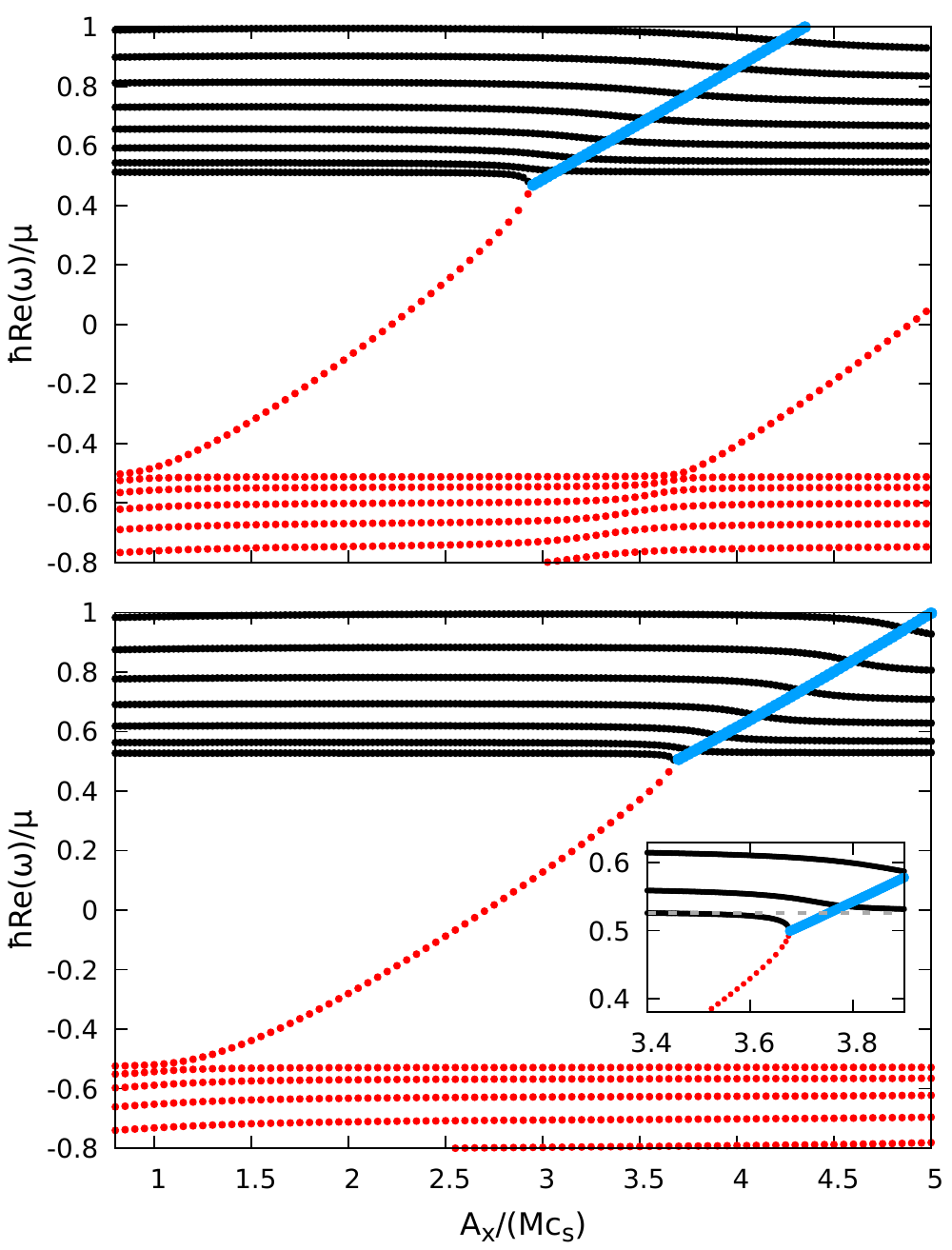}
  \caption{Spectra of the effective one-dimensional  Bogoliubov problem \eqref{eq:bogo-problem-gauge} for a condensate confined in a box $0\leq y\leq y_{max}$ in the presence of a transverse vector potential of variable intensity $A_x$ restricted to the region $y\in[0,L]$ with $L=2\xi$ and $y_{max}=30\xi$. The transverse momentum is fixed at $k_x=-0.5/\xi$ and the speed of sound $c_s$ is spatially uniform. Black solid (red dotted) lines indicate the real-valued frequencies of the positive (negative) norm modes; the blue thick lines are the real part of the frequencies of dynamically unstable zero-norm modes. The inset is a zoom of the region where dynamical instability emerges. The lower panel show the complete calculation, the upper part is the prediction of the hydrodynamic Klein-Gordon approximation as discussed in the Appendix.}
  \label{fig:ssw-inset}
\end{figure}

Here we consider the specific configuration that was addressed in Figure \ref{fig:gpe-evolution-noise} and we impose Dirichlet boundary conditions in $0$ and $y_{max}$ to the field $\vp$. The lower panel in Figure \ref{fig:ssw-inset} shows how the spectrum of this Bogoliubov problem varies as a function of the vector potential intensity $A_x$ for a fixed size $L$ of the moving region and a fixed transverse momentum $k_x$. In the electrostatic model, this corresponds to increasing the amplitude of the electrostatic potential $A_0$. One can see that at some point a negative norm state enters the mass gap: in the electrostatic case this corresponds to a bound antiparticle state localized in the positive electrostatic potential box. When the frequency of this state approaches the positive-norm band, opposite-norm states \textit{stick} together and give rise to zero-norm dynamically unstable modes (see Appendix) that can be thought as the continuous production of pairs of particles with opposite energies, one falling into the localized negative-norm mode, the other being radiated away on the positive-norm band. This is exactly the mechanism of the SSW effect.

\begin{figure}[t]
  \centering
  \includegraphics[width=\columnwidth]{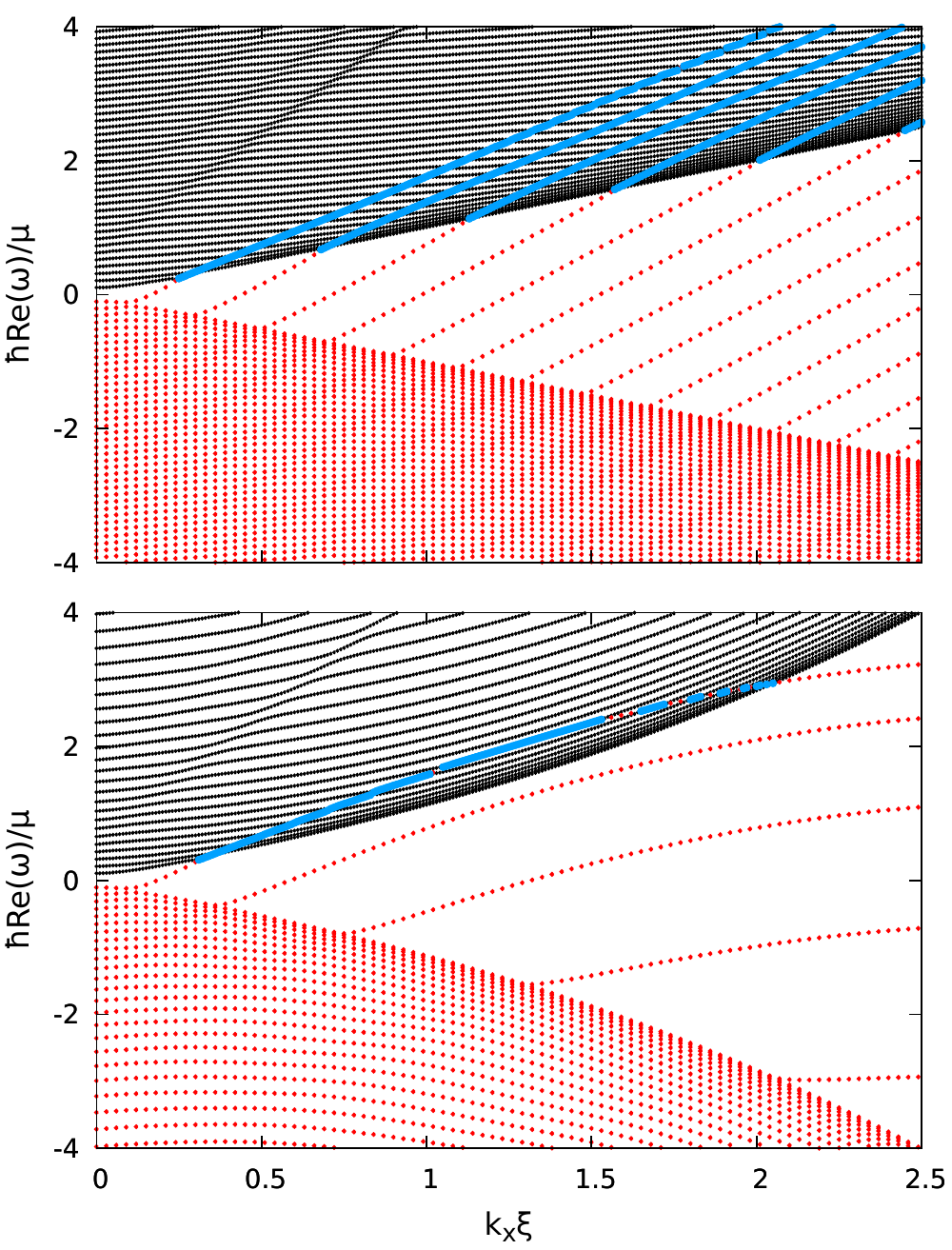}
  \caption{$k_x$-dependent spectra of the same effective one-dimensional Bogoliubov problem \eqref{eq:bogo-problem-gauge} as studied in  Figure \ref{fig:ssw-inset}. System parameters: $y_{max}=30\xi$, $L=4\xi$ and $A_x=-3Mc_s$. The upper panel shows the result of the hydrodynamical Klein-Gordon approximation, the lower panel one illustrates the complete Bogoliubov problem. The effects of superluminal dispersion are evident in the different spacing of the modes and in their curvature as a function of $k_x$, which result in the suppression of the instability at high transverse momenta.}
  \label{fig:ssw-dispersion}
\end{figure}

Along with the exact solution of the Bogoliubov problem that we just discussed, it is also interesting to consider this problem within the framework of the hydrodynamic approximation. This approximation corresponds in fact to the Klein--Gordon equation for which the SSW effect was originally derived. The result of such calculations (see the Appendix for the specification of the problem) is shown in the upper panel of Figure \ref{fig:ssw-inset}: except for some quantitative differences, the phenomenology is qualitatively identical.

The effects of superluminal nature of the Bogoliubov dispersion can be highlighted by performing an analogous calculation of the spectra as a function of the transverse momentum $k_x$ for fixed values of the size $L$ and of the vector potential intensity $A_x$. The results for both the exact problem and the hydrodynamic approximation are reported in the lower and upper panels of Figure \ref{fig:ssw-dispersion}, respectively. One can see that at small transverse momenta $k_x$ the behaviour is, as expected, essentially the same. In contrast, at large $k_x$ the presence of dispersion in the Bogoliubov problem has the consequence that both the mass gap and the bound states energy no longer show a linear dependence on $k_x$. In particular, for large enough $k_x$ the bound state reenters into the mass gap and the instability is correspondingly suppressed.

This impact of the superluminal dispersion onto the instability is very similar to the suppression of superradiant scattering for high transverse momenta. Also, condition \eqref{eq:velocity-condition} for superradiance is the same for the occurrence of instabilities. In the upper plot of Figure \ref{fig:ssw-dispersion}, this condition can be graphically understood in terms of the slope of the bound state. This slope is proportional to $|A_x|$ and for $|A_x|<2Mc_s$ is smaller than the slope of the $k_x$-dependent mass gap, that is of the lowest positive-norm state. As a result, instabilities can not develop in this case.

Our reasoning so far assumed open boundary conditions for large-$y$, so radiative waves can be emitted in this direction. Actually, the spectra shown in Figures \ref{fig:ssw-inset} and \ref{fig:ssw-dispersion} were calculated for finite size systems with Dirichlet boundary conditions. While the considered system size is generally large enough that the geometry can be generally considered as effectively open in the large-$y$ direction, following the arguments in~\cite{giacomelli2019ergoregion}, still some remnants of the finite size are visible in some specific parts of the spectra. For instance, a suppression of the instability is possible for specific parameter values: the dynamical instability is due to the coupling of two modes of opposite norm that are close to resonance. For a finite system the spectrum is discrete so that such pairs of modes may not be available. This is what happens in the lower panel of Figure \ref{fig:ssw-dispersion} around $k_x\xi\sim 2$, where the instability is absent for some intervals of the transverse momenta: even though the energetically unstable negative-norm mode is above the mass gap, it is far from resonance with positive-norm modes and the instability is effectively quenched. As it was discussed in full detail in~\cite{giacomelli2019ergoregion}, increasing the system size reduces the spacing between modes and removes the stability islands.

\subsection{Detection of dynamical instabilities via mode-matching}
\label{sec:modematch}

The discussion on dynamical instabilities carried out in the previus Subsections was based on a combination of numerical GPE simulations and a semi-analytical study of a standard Bogoliubov problem in a finite-size system with Dirichlet boundary conditions. In this Subsection, we introduce a variant of the Bogoliubov approach that naturally includes the open boundary conditions and is able to identify the intrinsic dynamical instabilities of an unbounded system without the need of artificially restricting to a finite size and then taking an infinite-size limit as discussed in~\cite{giacomelli2019ergoregion}.

The idea is the following. For a fixed (real) $k_x$ and different (complex) frequencies $\w$, we look for the roots of the dispersion relations \eqref{eq:bogo-dispersion} in each of the two uniform regions and we construct the associated plane-wave modes. Among all these modes, we focus on the ones that display an exponential decay away from the interface. The existence of global modes of the whole system at a given (complex) $\w$ satisfying the desired boundary conditions is then checked by trying to match the plane waves at the interface under the required continuity conditions. This imposes a linear set of equations to the mode amplitudes and the existence of non-trivial solutions at specific $\w$ is signalled by a vanishing determinant. This approach was used to prove the dynamical stability of a white-hole configuration in \cite{mayoral2011acoustic}.

One advantage of this method is that we can treat exactly open systems in the $y$ direction by selecting the appropriately decaying modes in each of the two regions. At fixed $k_x$ and $\w$ the Bogoliubov dispersion relation will in general have four $k_y$ roots. For non-real frequencies, all roots are complex, two with a positive imaginary part and two with a negative one. If open boundary conditions are considered, exponentially growing modes have to be discarded. This leaves us with only two relevant asymptotically bounded modes in each region, namely the ones with $\Im(k_y)>0$ in the faster upper region and the ones with $\Im(k_y)<0$ in the slower lower region.

In the fast/slow region the plane wave expansion of the modes will hence be
\begin{equation}
	\ket{\phi^{f/s}}=\sum_j\mathcal{A}^{f/s}_j
	\begin{pmatrix}
		1 \\ \b^{f/s}_j
	\end{pmatrix}
	 e^{ik^{f/s}_{y,j} y},
\end{equation}
where the sum runs over the physically relevant modes and $\beta_j^{f/s}$ is the proportionality constant between the two components of the Bogoliubov spinor that can be obtained from the linear problem \eqref{eq:bogo-problem-gauge} for a homogeneous system with the parameters of the fast/slow region.

\begin{figure}[t]
  \centering
  \includegraphics[width=\columnwidth]{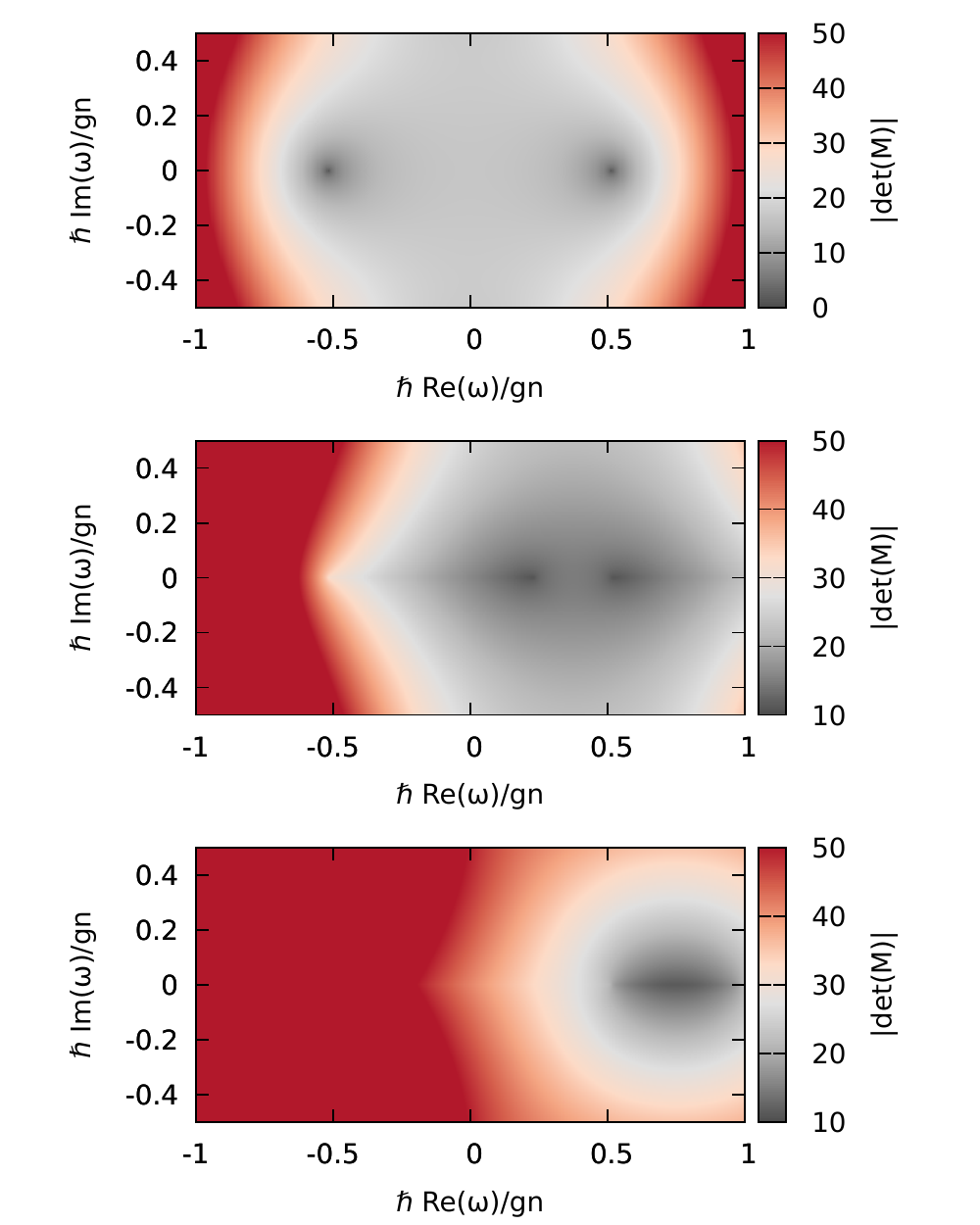}
  \caption{Log-scale color plots of the absolute value of the determinant (arbitrary units) of the linear matching problem with open boundary conditions on both sides of the interface for $k_x=0.5/\xi$. The three plots correspond to $A_x^f=0$, $A_x^f=-1.5\ Mc_s$ and $A_x^f=-3\ Mc_s$. A vanishing value of the determinant at a positive imaginary part of the frequency $\Im(\omega)$ is a necessary (but not sufficient) condition for instability. Here, zeros of the determinant are present only in the first (trivial) case and only for real values of the frequency, meaning that the single interface configuration is dynamically stable.}
  \label{fig:no-surface-instab}
\end{figure}

Consider first the configuration with open boundary conditions on both sides. In this case, we have two relevant modes per side. The continuity conditions at the interface for the two spinor components of the fluctuation field and of their first derivative give four linear conditions that can be used to determine the four mode amplitudes. The existence of non-trivial solutions requires for the determinant to be zero at some frequency $\w$. Dynamical instabilities are associated to roots with a positive imaginary part $\Im(\w)>0$.
Notice that this is a necessary but not sufficient condition for instabilities: zeros of the determinant may in fact also occur for frequencies for which the roots of the dispersion relation become degenerate within one of the two homogeneous regions. These zeros do not correspond to dynamical instabilities and can be easily identified by looking at the corresponding values of $k_y$. In our configuration, such zeros are found for $\w=\pm i\mu/\h$ and $\w=-A_x k_x/M\pm i\mu/\h$.

Beyond these spurious roots that must be discarded from the outset, in the case of a single interface no other solutions are found except for purely real frequencies in the case of a vanishing vector potential, 
as one can see in Figure~\ref{fig:no-surface-instab}. This proves that the condensate is dynamically stable and, in particular, does not show localized instabilities along the ergosurface, in stark contrast to classical hydrodynamic systems where analogous velocity fields are generally unstable against the generation of ripples at the interface.

\begin{figure}[t]
  \centering
  \includegraphics[width=\columnwidth]{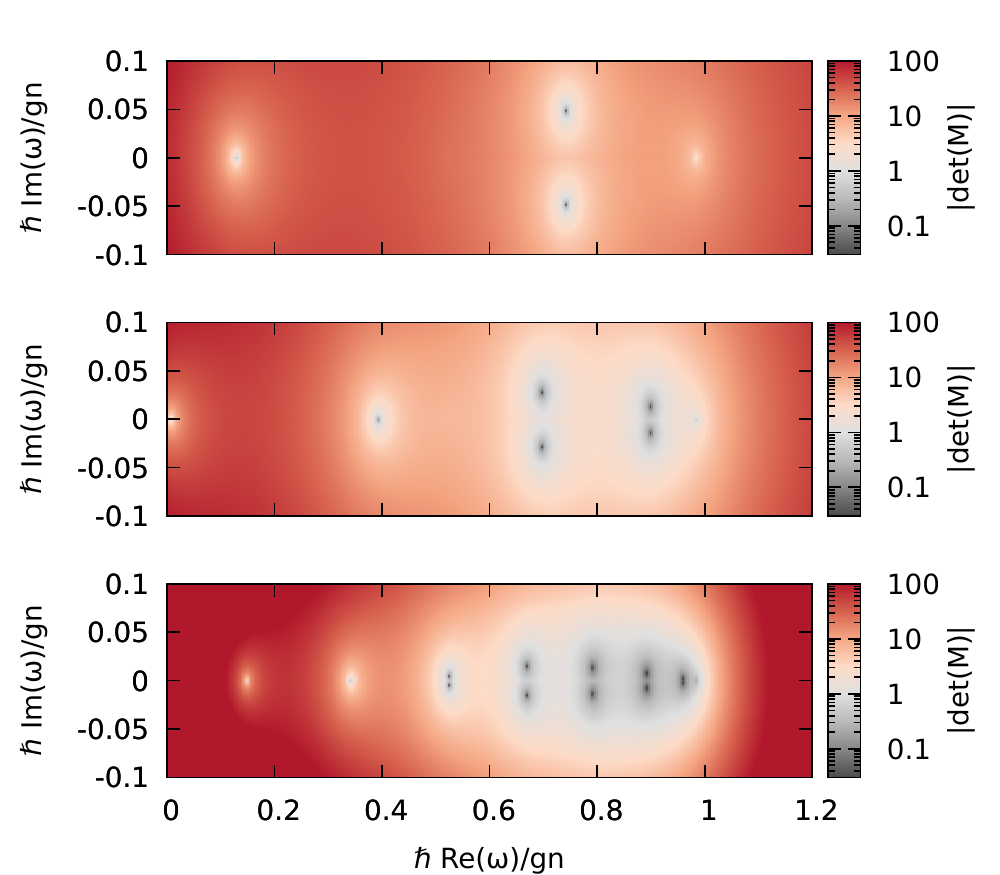}
  \caption{Log-scale plots of the absolute value of the determinant (arbitrary units) in the case of a reflecting boundary condition in the faster region at a distance $L$ from the interface. Here $k_x=0.5/\xi$ and $A_x^f=-3\ Mc_s$ and the tree plots correspond to $L=5\ \xi$, $L=10\ \xi$ and $L=20\ \xi$. Dynamical instabilities are signalled by the presence of roots in the upper half complex plane $\Im(\w)>0$. }
  \label{fig:det-SSW}
\end{figure}

Things of course change if we introduce reflecting boundary conditions on either side. Let us focus on the simplest case with a reflecting boundary condition in the fast region at a distance $L$ from the interface as discussed in the previous Section. In this case, we need to keep all the four roots of the dispersion relations in the upper region, corresponding to waves that propagate back and forth between the interface and the upper reflecting boundary. We then have six amplitudes to determine, the two extra conditions being provided by the condition that the field vanishes on the upper boundary. A plot of the resulting determinant in the complex-$\w$ plane is shown in Figure~\ref{fig:det-SSW}: as expected, the dynamical instabilities associated to the SSW effect discussed in the previous Section emerge as a series of zeros of the determinant in the $\Im(\w)>0$ half-plane. Their frequency $\Re(\w)$ lies within the superradiant frequency range, here located between $0.5\lesssim \h\w/gn\lesssim 1$.

As expected, the number of the unstable modes depends on the size of the faster region. The momenta of the trapped modes giving rise to instability must in fact satisfy a \textit{quantization} condition given by the finite size of the \textit{electrostatic} potential box. This is clearly visible in the growing number of zeros when increasing $L$ from the top to the bottom panel of Figure~\ref{fig:det-SSW}. 
On the other hand, the instability rate $\Im(\w)$ decreases while increasing the cavity size. Physically, this is also easily understood since amplification occurs upon bouncing on the ergosurface and the \textit{round trip time} of the excitations increases with $L$.

\subsection{Discussion}
\label{sec:discuss}

Based on our findings so far, let us summarize the connection between our predictions for flowing condensates and the dynamics of rotating spacetimes in gravitational physics. Consider for example a massless scalar field in the space-time surrounding an asymptotically flat Kerr black hole. The radial reduction of the problem for fields with a fixed azimuthal number is formally equivalent to our mapping to an electrostatic problem at fixed transverse momentum, except for the cylindrical instead of planar geometry. In the black hole case, the ergoregion is surrounded by an unbounded space on one side and by the black hole horizon on the other side. Both these regions then provide some effective \textit{dissipation} in the shape of open boundary conditions that exclude the backfeeding of radiation towards the ergosurface (further discussion on spacetimes with horizons will be given in the following Section). As a result, the Kerr black hole is dynamically stable against scalar field perturbations, even though it is energetically unstable since energy extraction is possible via superradiant scattering processes~\cite{brito2015superradiance}. In the analogy, this corresponds to a single ergosurface in an unbounded condensate on both sides along $y$, for which we have anticipated superradiant scattering and dynamical stability in Sec.\ref{sec:transverseflow}.

Removing (part of) the dissipation on one of the two sides gives rise to dynamical instabilities just as we have just seen for our analog model. If a strong enough reflection occurs inside the ergoregion (corresponding to a reflection at the upper edge of our setup), one has the so-called \textit{ergoregion instability}: in astrophysics, this can happen for \textit{ergostars} that have an ergoregion but no horizon and, in analogue models, for vortices with no drain \cite{oliveira2014ergoregion}. On the other hand, if the reflection occurs on the outside (corresponding to the lower part of our setup), the instability is known as a \textit{black-hole-bomb}: this happens for example for rotating black holes in asymptotically AdS spacetimes or for massive bosonic fields, whose low-frequency excitations (falling in the asymptotical mass gap) are naturally confined by the absence of propagating modes at infinity~\cite{brito2015superradiance}.

Whereas this discussion confirms the usual picture of superradiant phenomena, it is important to make some general comments on the actual role of open boundary conditions and dissipation mechanisms in superradiance. It is sometimes stated (see for example \cite{richartz2009generalized}) that no amplified scattering can be observed in the absence of efficient dissipation channels able to evacuate the amplified waves, e.g. through the horizon. Our discussion in this Section and the calculations performed in~\cite{giacomelli2019ergoregion} for quantized vortex geometries have shown that the situation is slightly more complex than that, as was already pointed out in \cite{vicente2018penrose}. On one hand, dynamical instabilities do indeed emerge in the absence of dissipation preventing amplified scattering of incident plane waves that extend indefinitely over time.
On the other hand, the plots on the upper row of Fig.\ref{fig:gpe-evolution-noise} show that finite-width wavepackets can be significantly amplified upon reflection on the ergosurface, even if this eventually ends up triggering some instability at later times. Amplification is in fact due to the coupling between opposite-norm modes in a restricted region of space around the interface. If this process is well separated in space from the reflecting element, the positive feedback mechanism responsible for the instability only occurs after a sizable time-interval set by the round-trip time of the transmitted wavepacket. In the meanwhile, only the amplified wavepacket is visible. As a result, while the \textit{time-independent} spectrum of the system does show clear signatures of dynamical instabilities, superradiance can still be observed as a \textit{time-dependent} process.

\section{The effect of horizons}\label{sec:horizon}

A common thread of the previous Sections has been that an efficient amplified reflection requires some effective dissipation mechanism inside the ergoregion to evacuate the product of superradiant amplification and avoid dynamical instabilities due to the repeated amplification of the negative-energy waves. In the classical treatments of superradiance it is pointed out how such a  mechanism is naturally provided in black hole spacetimes by the horizon, which acts as an open boundary condition, prohibiting the reflection of radiation towards the ergosurface. 
In this Section, we show how the behaviour can be much richer than this, in particular we will present configurations where ergoregion instabilities may occur despite the presence of a horizon.

\subsection{Scattering at a horizon}
\label{sec:horizon_scatt}

\begin{figure}[t]
  \centering
  \includegraphics[width=\columnwidth]{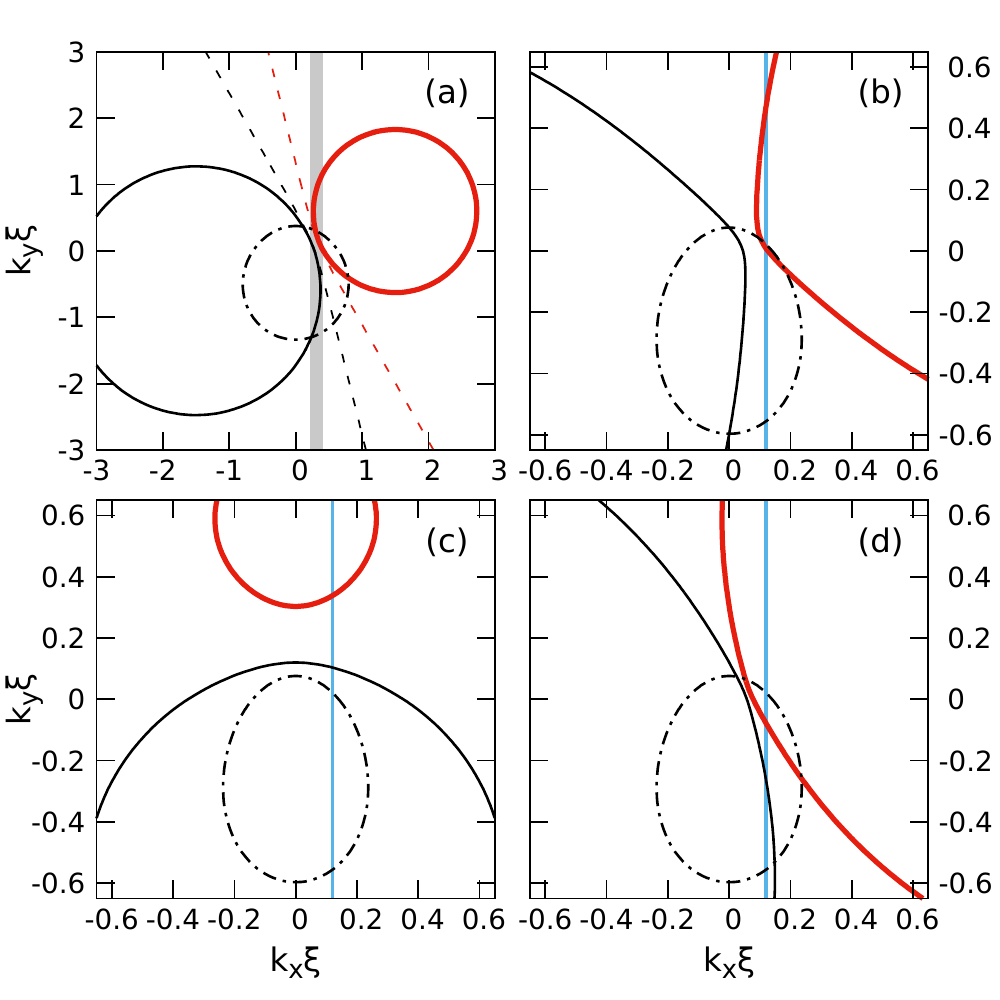}
  \caption{(a) Comparison of the cuts at $\hbar\w/Mc_1^2=1$ of the Bogoliubov dispersion relations inside (solid lines) and outside (dash-dotted lines) the \textit{horizon} for $v^s_x=0$, $v^f_x=2.12Mc_1$,  $v^{s}_y=v^f_y=0.85c_1$ and $c_2=0.3c_1$. Dashed lines are instead the cut of dispersion relation of the corresponding Klein-Gordon problem inside the horizon. (b)-(d) Referring to the configuration of Figure \ref{fig:withorizon-section}: comparison of the  $\hbar\w/Mc_1^2=0.2$ cut of the Bogoliubov dispersion relations in the exterior region (dash-dotted lines) with the one in the ergoregion (b) and with the ones inside a region of supersonic longitudinal flow $v_y^f>c_2$ (c,d). Parameters: $A_E=-2.12Mc_1$, $v^{s}_y=v^f_y=0.85c_1$, $c_2=0.3 c_1$. In panel (c), $A_H=0$. In panel (d), $A_H=A_E$. The vertical blue lines in (b-d) indicate the value $k_x\xi=0.1$ considered in the following Figs.\ref{fig:withorizon-section} and \ref{fig:withorizon-evol}. }
  \label{fig:withorizon-cases}
\end{figure}

As a first step, consider an interface at $y=0$ separating a slow region of subsonic flow $v^s<c_1$ in the lower half-plane $y<0$ from a fast region of supersonic flow $v^f>c_2$. In contrast to the previous sections, the flow velocities $\mathbf{v}^{s,f}$ are no longer assumed to be oriented along the direction $x$ parallel to the interface, but can have different directions, and we do not assume that the speed of sound is spatially uniform, in particular we focus on the $c_1>c_2$ case.

Let us start from the hydrodynamic Klein-Gordon regime. For generic flow velocity directions, the dispersion of the modes in each region can be obtained by rotating the curves in Fig.\ref{fig:kg-dispersion} in the $(k_x,k_y)$ plane. In the upper region of supersonic $v^f>c_2$ flow, the asymptotes of the hyperbolas are oriented in a different way depending on whether the $y$ component of the velocity $v^f_y$ is smaller or larger than the speed of sound $c_2$, which gives rise to different scattering processes.
In the former case, one has a slight rotation of Fig.\ref{fig:kg-dispersion}, namely there still exist two windows of $k_x$ values in which one only has a positive or a negative norm mode and these two regions are separated by an interval with no available mode. As a result, the same superradiance physics takes place: depending on $k_x$, an incident wavepacket coming from the subsonic region at $y<0$ can either be totally reflected (region II in Fig.\ref{fig:kg-dispersion}, or be partially transmitted and reflected (regions III and IV), or undergo superradiant scattering (region I). 

In the latter $v^f_y>c_2$ case, the orientation of the asymptotes is the one displayed by the dashed lines in Fig.\ref{fig:withorizon-cases}(a). As expected for a horizon, all the modes have a positive $y$-component of the group velocity, so they cannot travel back through the horizon. In particular, both a positive and a negative norm mode are available for any value of $k_x$:
as a result the incident wavepacket will split in a pair of transmitted and reflected positive-norm components in addition to the negative-norm transmitted one. In spite of the amplification given by the negative-norm mode, because of this multi-partite splitting, the intensities of the wavepackets are not necessarily larger than the incident one and superradiance in the sense of amplified reflection does not generically occur. 

The situation changes dramatically when the superluminal Bogoliubov dispersion is considered. In this case, illustrated by the solid lines in Fig.\ref{fig:withorizon-cases}(a), there exist again regions of $k_x$ values in which a single negative norm mode is available, which leads to the unexpected behaviour of a superradiant scattering occurring directly at the horizon in the absence of an isolated ergosurface. GPE simulations of the wavepacket dynamics (not shown) give results qualitatively identical to the ones in Sec.\ref{sec:transverseflow}. This reflects the fact that horizons are ill-defined for a field of superluminal dispersion relation. However, for transverse momenta in the gray region of Fig.\ref{fig:withorizon-cases}(a), positive- and negative-norm modes coexist in the faster region, so that, despite superluminality, it behaves for large-wavelength modes as the interior of an horizon.

Solid lines in panel (c) of the same figure show instead an analogous dispersion for $v^f_x=0$ case: in this case, there are no $k_x$ values for which negative-norm modes only exist, so no purely superradiant scattering is possible. This shows how a lateral flow is anyway an essential ingredient of superradiance.

\subsection{A toy model for rotating black holes}
\label{sec:toy_bh}

\begin{figure}[t]
  \centering
  \includegraphics[width=\columnwidth]{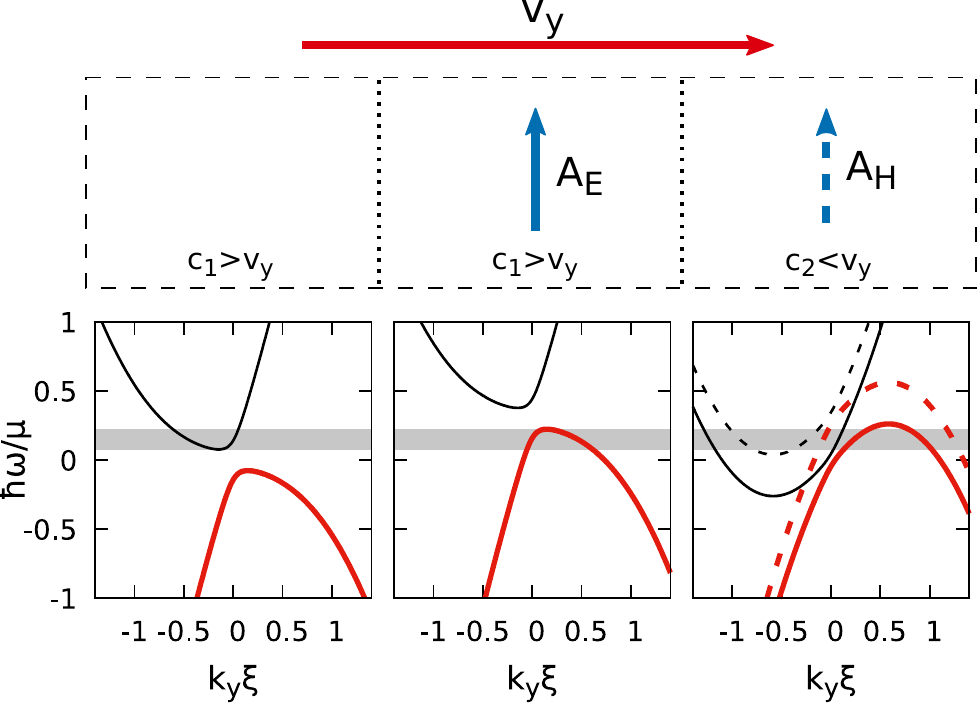}
  \caption{Top row: scheme of a configuration including both an ergoregion and an horizon. The ergoregion is included by means of a vector potential oriented along $x$. The horizon is created by changing the speed of sound in the third region. Here, the vector potential $A_H$ may or may not be present. Bottom row: plots of the dispersion relations at fixed transverse momentum $k_x\xi=0.1$ for the same parameters considered in Fig.\ref{fig:withorizon-cases}(a-c), namely $A_E=2.12Mc_1$, $v_y=0.85c_1$, $c_2=0.3 c_1$, $A_H=A_E$ (dashed) or $0$ (solid). The gray region indicates the frequency interval in which superradiant scattering can occur.
  }
  \label{fig:withorizon-section}
\end{figure}

Inspired by general-relativistic black holes, let us now focus on configurations displaying an external ergosurface and an internal horizon like the one sketched in the upper part of Figure \ref{fig:withorizon-section}. This configuration consists of three layers and displays a finite longitudinal velocity along $y$ in addition to the synthetic vector potential directed along $x$. In the outermost layer (left), the speed of sound $c_1$ exceeds all components of the velocity and the flow is subsonic. In the central layer, the vector potential $A_E$ is large enough to give a supersonic flow along $x$, but the inward radial velocity $v_y$ is still subsonic. Except for the small longitudinal speed, the first interface is expected to behave very similarly to the ergosurface discussed in the previous Sections. In the third layer (right), the longitudinal velocity $v_y$ exceeds the speed of sound $c_2$, so the second interface behaves as a horizon for long-wavelength waves. The synthetic vector potential $A_H$ in the third region is drawn with a dashed line to indicate that we are going to consider values ranging from $A_H=A_E$ (which resembles a vortex with drain) to $A_H=0$.

In Figure \ref{fig:withorizon-cases} we show comparisons between the fixed-$\w$ cuts of the Bogoliubov dispersion relations in the left, subsonic region (dash-dotted lines) and in the supersonic central and right regions (solid lines). Panel (b) shows the comparison between the left and the central ergoregion: the flow along $y$ is responsible for a \textit{tilt} of the curves in the ergoregion with respect to the $v_y=0$ case shown in Figure \ref{fig:bogo-cases}(d) but the structure of the available modes remains essentially the same. Panels (c) and (d) show instead the comparison with the third region inside the horizon, in respectively the $A_H=0$ and $A_H=A_E$ cases.

The vertical blue lines in the panels (b-d) indicate a value of the transverse wavevector $k_x\xi=0.1$ at which, for the chosen frequency value $\hbar\omega/Mc_1^2=0.2$, the central region behaves as an ergoregion and the right one as the interior of a black hole. Focusing on this specific value of $k_x$, in the lower part of Figure \ref{fig:withorizon-section} we show constant-$k_x$ cuts of the dispersion relations. The choice of parameters is such that, for all the values between $A_H=0$ and $A_H=A_E$ of the vector potential in the third region, the interface between the second and third regions behaves as an horizon (that is inside the horizon both positive- and negative-norm modes are available at the same frequency) at all frequencies for which amplified reflection at the ergosurface is possible (gray region).

Based on general arguments of black hole physics \cite{bardeen1972rotating,berti2009quasinormal}, we could expect that the horizon behaves as a perfectly absorbing element, so that all the negative-norm excitations created by superradiant processes at the ergosurface are dumped into the black hole. In fact, Kerr black holes in asymptotically flat spacetimes are found to be dynamically stable with respect to massless scalar field perturbations \cite{brito2015superradiance} thanks to the presence of an horizon that provides an open boundary condition and to the negligible amount of reflection by the gravitational potential barrier. However, from a different perspective, moving away from general relativistic black holes one could also expect that the presence of an horizon is not enough to assure dynamical stability since a sizable reflection inside the ergoregion may well occur.

This conjecture can be tested by studying the one-dimensional Bogoliubov--de Gennes equation \eqref{eq:bogo-problem-gauge} in the configuration sketched in Fig.\ref{fig:withorizon-section}. As done in~\cite{balbinot2008nonlocal,carusotto2008numerical,recati2009bogoliubov}, a spatially varying speed of sound is obtained by means of a suitable spatial profile of the interaction constant and of the external potential, taken as specified in Table \ref{tab:spacedependence}, so to maintain a constant density for the background condensate $\Psi_0(y)=\sqrt{n_0}e^{i(Mv_y/\h)y}$. As usual, the interaction constant is related to the speed of sound via $c_{1,2}=\sqrt{g_{1,2}n_0/M}$ and, with the parameters specified in Figure \ref{fig:withorizon-section}, we have the configuration summarized in Table \ref{tab:spacedependence}.

\begin{table}[b]
\begin{tabular}{c||c|c|c}
				& $A_x(y)$ & $g(y)$ &  $V(y)$   \\
	\hline
	\hline
 $y<y_E$		& $0 $	& $g_1$ &  $0$   \\
 $y_E<y<y_H$ 	& $A_E$	& $g_1$ &  $\displaystyle -\frac{A_E^2}{2M}$  \\
 $y>y_H$		& $A_H$	& $g_2$ &  $\displaystyle -\frac{A_H^2}{2M}+(g_1-g_2)n_0$   
\end{tabular}
\caption{Spatial dependence of the vector potential field, of the interaction constant and of the external potential to obtain the configuration of Fig.\ref{fig:withorizon-section}.}
\label{tab:spacedependence}
\end{table}

Instead of taking sharp jumps of these quantities in passing between the regions, we consider smooth spatial spatial variations of the following forms for the interaction constant
\begin{equation}\label{eq:smooth-horizon}
	g(y)=g_1+\frac{g_2-g_1}{2}\left[1+\tanh\left(\frac{y-y_H}{\ell_H}\right)\right]
\end{equation}
and for the vector potential
\begin{equation}
\begin{split}
	A_x(y)=&\frac{A_E}{2}\left[1+\tanh\left(\frac{y-y_E}{\ell_E}\right)\right]\\ &+\frac{A_H-A_E}{2}\left[1+\tanh\left(\frac{y-y_H}{\ell_H}\right)\right]\,.
	\end{split}
\end{equation}
Here, $y_E$ and $y_H$ are the points around which the transitions to the ergoregion and to the black hole interior are centered, and $\ell_E$ and $\ell_H$ regulate the smoothness of these transitions.

\subsection{Ergoregion instabilities in the presence of an horizon}
\label{sec:horizon_dyn}

To detect the presence of dynamical instabilities we perform numerical integrations of the time-dependent Bogoliubov problem \eqref{eq:bogo-problem-gauge} in the configuration described above, starting from an initial noisy configuration to offer a seed to the unstable modes. A pair of absorbing regions are included well outside the ergoregion and well inside the horizon so to mimic open boundary conditions. This guarantees that all spurious instabilities that may come from the backfeeding of excitations from the outside (black hole bomb) and from the inside (black hole lasing) of the black hole are fully suppressed. 

Even if we saw that the structure of the modes in the three regions is the same for all the values between $A_H=0$ and $A_H=A_E$ of the vector potential inside the horizon, the numerical results turn out to be qualitatively different depending on the difference between $A_H$ and $A_E$. Moreover, while the thickness $\ell_E$ of the transition at the ergoregion does not alter the physics substantially (and one can also consider a sharp jump as in the rest of the paper), the \textit{smoothness} $\ell_H$ of the transition at the horizon can determine largely different behaviours.

In the presence of a second jump of vector potential from $A_x=A_E$ to $A_x=A_H=0$, for very sharp horizons (that is for small $\ell_H$) dynamically unstable modes localized on the interface and independent on the size of the ergoregion are observed; these do not seem to be directly related to superradiant phenomena but depend on the microscopic physics of the condensate. This will not be further discussed here and will be addressed in future work.

\begin{figure}[t]
  \centering
  \includegraphics[width=\columnwidth]{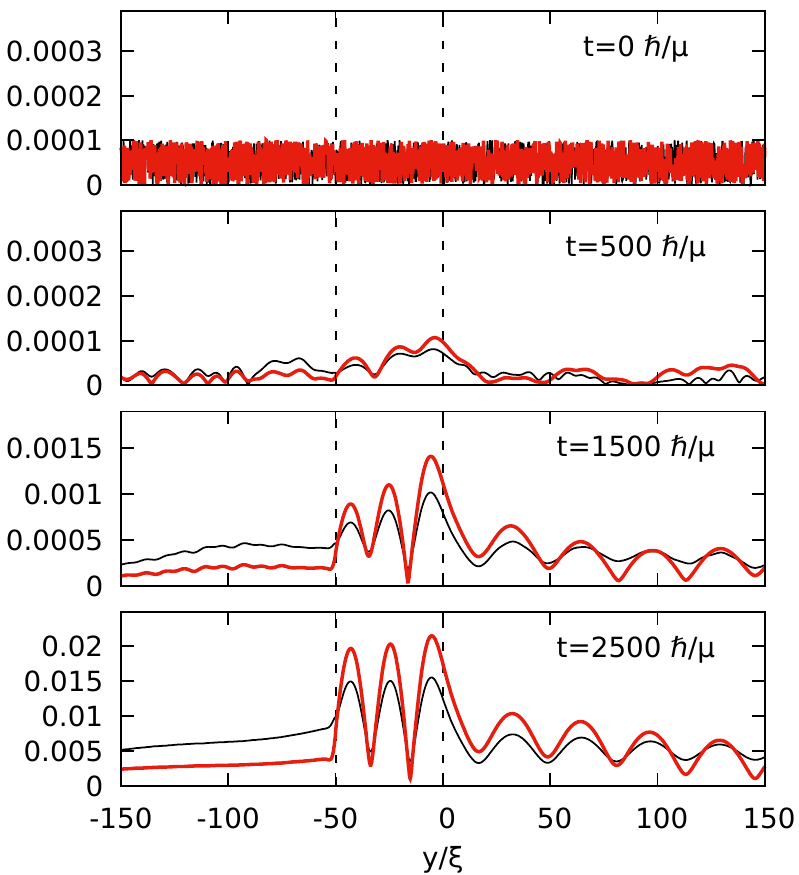}
  \caption{Snapshots of the time evolution of the moduli of the fluctuation spinor components given by the reduced one-dimensional Bogoliubov problem at a fixed transverse wavevector $k_x\xi=0.1$ for the configuration sketched in Figure \ref{fig:withorizon-section} with $A_H=0$ and $\ell_0=5\xi$. The evolution starts from a noisy configuration and absorbing regions are included at the boundary of the integration box to avoid spurious reflections. The black (thinner) and red (thicker) lines show the modulus of the $u,v$ components of the Bogoliubov spinor. Time is measured in units of the external $\mu=mc_1^2$. 
  }
  \label{fig:withorizon-evol}
\end{figure}

For smoother horizons, such localized instabilities are no longer present, and a spatially extended dynamical instability of completely different nature takes place, as signalled by the fast temporal growth of a spatially oscillating pattern between the ergosurface and the horizon. An example of such a temporal evolution is shown in Figure \ref{fig:withorizon-evol}. The origin of this instability can be traced to the self-amplification of the excitations trapped in the ergoregion according to a mechanism similar to one one illustrated in Fig.\ref{fig:gpe-evolution}: excitations bounce between the ergosurface, where they get amplified, and the transition region of the horizon, where they are reflected with a sufficiently high amplitude to give an overall increase of the excitation intensity during a round-trip. These are the typical features of an \textit{ergoregion instability}: interestingly enough, such instability occurs here in spite of the presence of a horizon. Note that here the radiative wave that enters the horizon is composed of two $k_y$ components of opposite norms whose beating is responsible for the oscillating behaviour seen in the figure for $y>0$. 

Further confirmation of this physical picture is provided by a spectral analysis of the growing perturbation. This shows that the frequencies of all unstable modes fall in the superradiant interval (gray region of Figure \ref{fig:withorizon-section}). Moreover, we checked that increasing the size of the ergoregion causes the number of unstable modes to increase and their instability rates to decrease according to the increased round-trip time within the ergoregion.

While this instability may be ascribed to a reflection given by the relatively sharp transition between the second and third regions, that is to the deviation from the hydrodynamic regime, a further increase of $\ell_H$ shows that, even if the growth rate decreases, the instability does not disappear even for very smooth transitions ($\ell_H\sim 100\xi$).
The situation is different if we reduce the second jump of the vector potential $A_H-A_E$. We performed simulations for $A_H$ ranging from zero to a flat vector potential profile across the horizon interface $A_H=A_E$, finding that no instability is present for high enough $A_H$. The initial noisy configuration smoothly decays in time and only displays some \textit{quasinormal modes}. This configuration is thus close to the standard astrophysical case where reflection of waves traveling towards the black hole horizon is typically not present.

The difference between the two behaviours can be understood with a simple analysis in the spirit of the WKB approximation, as done for example in \cite{patrick2020superradiance, patrick2020rotational}. If the background spacetime varies over length scales much larger than the wavelength of the excitations, one can understand the propagation of fields by considering at each point the dispersion relation of a uniform medium depending on the local properties of the spacetime. In this picture, excitations of frequency $\w$ will  \textit{adiabatically} follow the local dispersion relation of the given branch, until they encounter a \textit{turning point} where propagating states cease being available and the frequency $\w$ enters into a gap of the local dispersion relation.
At these points the WKB approximation breaks down, the wavevector turns complex, and energy exchange between modes is possible. 
In our case, the points where this happens can be understood by following, at the given value of the conserved $k_x$, the spatial dependence of the lower edge of the upper positive-norm branch of the dispersion relation, $\mathrm{min}_{+}(y) := \min_{k_y}[\w_+(y,k_x,k_y)]$, and of the upper edge of the lower negative-norm branch, $\mathrm{max}_{-}(y) := \max_{k_y}[\w_-(y,k_x,k_y)]$. The turning point occurs at the position $y_{tp}$ where the frequency $\w$ hits the edge of its branch, e.g. $\w=\mathrm{min}_+(y_{tp})$ ($\w=\mathrm{max}_-(y_{tp})$).
As compared to analogous WKB treatments of wave reflection processes in solid-state physics~\cite{ashcroft1976solid} and optics of light~\cite{burstein2012confined} and matter~\cite{carusotto2000modulated} waves, our superradiant effects originate from the fact that the two branches display opposite signs of the conserved norm.

\begin{figure}[t]
  \centering
  \includegraphics[width=\columnwidth]{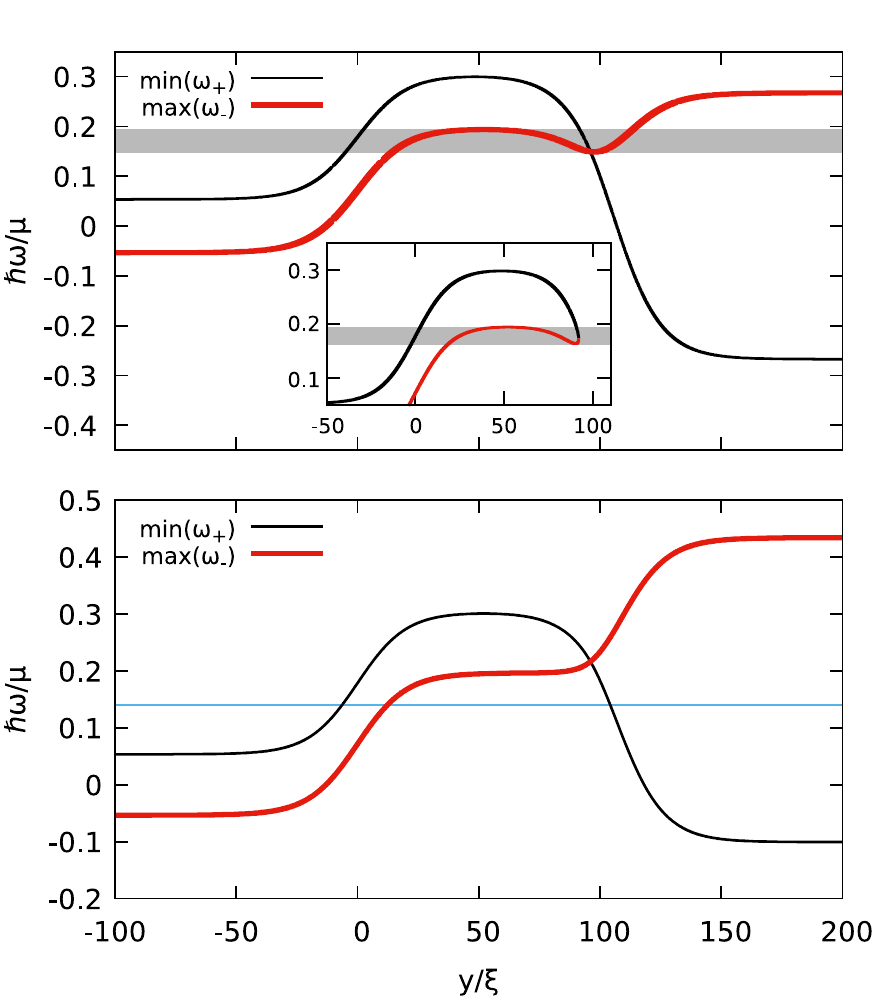}
  \caption{Space dependence of the minimum frequency of the upper branch of the dispersion relation $\textrm{min}_+(y)$ (black thin lines) and of the maximum of the lower dispersion branch $\textrm{max}_-(y)$ (red thick lines) in two configurations realizing the sketch in Fig.\ref{fig:withorizon-section}. Both panels refer to fixed $k_x\xi=0.1$, $c_2=0.3\;c_1$, $v_y=0.85\;c_1$, $A_E=2.12\;Mc_1$ and the thickness of the two transition regions are $\ell_E=\ell_H=20\xi$. The upper panel corresponds to the case $A_H=0$, with the gray region indicating the frequency range in which ergoregion instabilities occur. In the inset the corresponding non-dispersive Klein--Gordon case. The lower panel corresponds to $A_H=(2/3)A_E$, for which no ergoregion instabilities occur; the horizontal blue line indicates a generic frequency at which stable superradiant scattering occurs.}
  \label{fig:withorizon-wkb}
\end{figure}

Following this picture, analogously to what was done in \cite{patrick2020rotational}, in Figure \ref{fig:withorizon-wkb} we plot the edges of the two branches as a function of the position $y$ for two choices of $A_H$ leading to the two different behaviours mentioned above.
Let us start from the case of a relatively large $A_H$ displayed in the lower panel. Consider a positive-norm wave coming from negative $y$ at the frequency indicated by the horizontal blue line. When this line intersects the black one, no propagating mode is any longer available at its frequency and the wave gets reflected. Since states on the lower dispersion branch are available at larger $y$, partial tunneling of the wave onto this branch is possible. Since the negative norm of the tunneled wave must be compensated by amplified reflection on the positive-norm branch, superradiance naturally occurs. Since no more turning points are present on the negative branch, the tunneled wave freely propagates towards large $y$ suppressing the possibility of ergoregion instabilities.

The situation is different in the $A_H=0$ case displayed in the upper panel. Here, additional turning points are present on the lower branch. In particular, for frequencies in the gray region, a second pair of turning points is available around $y=100\;\xi$, so that the negative-norm wave can undergo reflection. Together with the superradiant process taking place around $y=15\;\xi$ for waves moving in the opposite direction, these repeated amplified reflections lead to an effective confinement of the negative-norm mode inside the ergoregion associated with a growth of its amplitude, leading to the ergoregion instability.

It is interesting to comment that stationary points of the curves of Fig. \ref{fig:withorizon-wkb} are associated to the existence of \textit{orbits}, that is trajectories moving along $x$ with vanishing velocity along $y$; these are the equivalent of \textit{light rings} in circular geometries. Maxima of the black thin curves $\mathrm{min}_+(y)$ and minima of the red thick curves $\mathrm{max}_-(y)$ correspond to the locations of unstable orbits that can radiate to the surrounding propagating modes of the same band. Minima of the black curves and maxima of the red curves correspond instead to stable orbits, since fluctuations living there can radiate away only through tunneling. These stable orbits, when occurring inside the ergoregion, give rise to the structure of turning points that we just discussed and are hence associated to the presence of ergoregion instabilities.

It is important to notice that this kind of ergoregion instability is not restricted to the superluminal dispersive case we are considering. An analogous analysis of the Klein--Gordon dispersion throughout our toy model is displayed in the inset of the upper panel of Fig.\ref{fig:withorizon-wkb}. This differs from the dispersive case for the absence of extrema of the dispersion branches inside the horizon. However the same structure of turning points outside the horizon is present, so that again ergoregion instabilities occur in the frequency range indicated by the gray region.

On the basis of this analysis, we can conclude that the presence of an horizon does not \textit{in general} guarantee the absence of ergoregion instabilities. Observable ergoregion instabilities may in fact arise for scattering on the gravitational potential past the ergosurface. The fact that no such instabilities have been ever found in general-relativistic black holes \cite{brito2015superradiance} or in draining vortex geometries \cite{berti2004quasinormal} is associated to the fact that these spacetimes do not display stable negative-energy orbits, that instead can exist around ultracompact objects with no horizon \cite{cardoso2014light,cunha2017light}. Still, as our calculations show, this feature does not appear to be a general property of spacetimes with an horizon and may not be satisfied in other configurations.

\section{Conclusions and discussion}
\label{sec:conclusions}

In this work we have proposed the use of synthetic vector potential fields in Bose--Einstein condensates as a new analog model platform where to quantum simulate a large variety of quantum field theories in curved spacetimes. As compared to standard superfluids, the application of a synthetic vector potential removes in fact the irrotationality constraint and allows to realize a wider range of flow profiles. In turn, this widens the spectrum of spacetime geometries that can be studied in the framework of analog models. 

Here, we have made use of this new tool to theoretically investigate the fundamental mechanisms of superradiance without being limited to the traditional vortex geometry. This has allowed us to disentangle the different effects at play in superradiant phenomena and highlight their rich interplay with black hole horizons. This provides new intuitive insight on some delicate issues of superradiant phenomena and clarifies some conceptual  debates in the literature.

In simplest terms, superradiant scattering can be understood as a mode-mixing scattering process taking place at the ergosurface, where an incident positive-energy wave gets amplified via transmission of a negative-energy wave inside the ergoregion. 
While in rotating black holes the conserved energy can be negative inside the ergoregion because of the time-time component of the metric changing sign,  negative-energy modes naturally appear in moving condensate as soon as the flow velocity exceeds the speed of sound. A graphical construction based on the Bogoliubov mode dispersion is proposed to determine and characterize the different scattering processes that can take place in each configuration, including negative refraction and superradiant reflection.

The fact that superradiance is based on a spatially localized mode-mixing process implies that the presence of dissipation is not a necessary ingredient for amplified scattering to occur: a wavepacket impinging on the ergosurface will in fact undergo amplification regardless of the presence of additional reflecting or absorbing elements in the neighborhood of the ergosurface and of the global dynamical stability of the configuration. Nevertheless, dissipation is a fundamental ingredient to avoid all those dynamical instabilities that emerge at later times when the amplified wave is fed back to the amplifying element. This can occur when reflecting boundary conditions are imposed, or when several interfaces are present. 

While open boundary conditions or spatially unbounded geometries provide a natural mechanism for evacuating the amplified waves and suppressing such instability mechanisms, we have found that the presence of an horizon is not always able to remove instabilities. Depending on the specific geometry, back-scattering of waves traveling towards a black hole horizon may in fact turn the superradiant scattering into a dynamical instability and therefore destabilize the ergoregion. The fact that this does not occur in general-relativistic black holes is to be ascribed to the specific properties of the spacetimes and is not a general feature of spacetimes with an ergoregion and a horizon.

While our discussion has been carried out with analog models of gravitational physics in mind, a similar phenomenology is also found for massive charged scalar fields. When such a field hits a jump in the electrostatic potential, amplified reflection processes may occur in the so-called bosonic Klein paradox. In more complex geometries, dynamical instabilities have been anticipated to occur, e.g. in the presence of square box electrostatic potentials, the so-called Schiff-Snyder-Weinberg effect. Introducing a synthetic vector potential in our analog model thus provides a flexible platform where to quantum simulate this physics in different geometries.

As already mentioned, our work does not mean to  quantitatively reproduce any specific system of general relativity. We rather aim at providing a general conceptual framework where known results can be inserted and understood under a different light and which can offer an intuitive guidance in further studies of black hole physics. Moreover, our work can be of help in the study of dispersive fields in black hole spacetimes. This problem is of great interest in the general relativity context in the light of understanding the effects of a possible breaking of the Lorentz symmetry at small scales~\cite{jacobson2006lorentz} and has already led to the remarkable discovery of the black hole lasing effect in the presence of an inner horizon~\cite{corley1999black}.

So far, our work has focused on the dynamics in a small excitation regime, where the linearized Bogoliubov approximation is an accurate description of the dynamics. This is valid for the superradiant scattering of weak incident wavepackets and accurately describes the early stages of the dynamical instability. A natural next step is to extend our investigation to the case of large amplitude perturbations for which the nonlinear coupling between Bogoliubov modes is important. In spite of the obviously different form of the nonlinear coupling in the Gross-Pitaevskii equation for condensates and in the Einstein equations for gravity, this problem is of high potential interest for the gravitational community as it provides a {\em toy model} for the back-reaction effect of the superradiant instability on the underlying metric, e.g. of rotating or charged black holes in anti-de Sitter space-times~\cite{PhysRevLett.116.141102,PhysRevLett.116.141101,PhysRevLett.119.041101}. First steps in this direction has been made in the context of the late-time evolution of black hole lasers based on atomic condensates~\cite{Michel_2013,Michel_2015,Nova_BHLaser1,Nova_BHLaser2}.

Along a related direction, our study of the scattering of classical waves and on the development of dynamical instabilities leading to macroscopic excitations is the natural starting point to attack quantum features of the field theory on curved space-time. Given the same mode-mixing origin of superradiant and Hawking processes~\cite{brito2015superradiance,carusotto2008numerical,recati2009bogoliubov}, it is natural to expect that superradiance in condensates will lead to the spontaneous production of phonon pairs with specific signatures in the correlation functions. The long-term perspective will be to explore the interplay between the quantum spontaneous superradiance and the backreaction effects and identify novel paths to black hole evaporation. 

From the experimental point of view, even though our discussion has been carried out with an special eye to atomic condensates, the general ideas developed in this work are of direct applicability also to analog models based on quantum fluids of light~\cite{RMP_2013,Jacquet,nguyen2015} for which synthetic magnetic fields are currently under active study~\cite{RMP_2019}.

\section{Acknowledgements}
\label{sec:Acknowledgements}

This work received funding from the European Union Horizon 2020 research and innovation program under grant agreement No.820392 (PhoQuS), the Provincia Autonoma di Trento and the Q@TN initiative. Stimulating discussions with Massimiliano Rinaldi, Nicolas Pavloff, Stefano Vitale, and our forever missed friend and master Renaud Parentani are warmly acknowledged.

\appendix*

\section{Norm, energy and the Bogoliubov and Klein--Gordon problems}
In Section \ref{sec:metric} we reported the hydrodynamic shape of the Bogoliubov--de Gennes equations for linear perturbations around some stationary state of the Gross--Pitaevskii equation. In this representation one derives the Klein--Gordon equation for phase fluctuations once the hydrodynamic approximation is taken.

However we analyzed superradiant amplification and instabilities using the concepts of energy and norm of the modes that are more natural for the Bogoliubov problem than for the Klein--Gordon one. Here, for completeness, we summarize some different formulations of the Klein--Gordon and the Bogoliubov problems, the connection between them and the properties of the eigenmodes.

By linearizing the GPE around some stationary state $\Psi_0$ as 
\begin{equation}\label{eq:linearization}
	\Psi=\Psi_0(1+\vp)=\sqrt{n}e^{i\Theta}\left(1+\frac{\delta n}{2n}+i\delta\Theta\right),
\end{equation}
one obtains the Bogoliubov-de Gennes equations that can be written in the convenient spinorial shape
\begin{equation}\label{eq:bogoliubov-matrix}
	i\h\de_t\begin{pmatrix} u_\vp\\ v_\vp \end{pmatrix} =
	\begin{bmatrix}
		D+gn & gn\\
		-gn & -D^*-gn
	\end{bmatrix}
	\begin{pmatrix} u_\vp\\ v_\vp \end{pmatrix}
	=\L \begin{pmatrix} u_\vp\\ v_\vp \end{pmatrix},
\end{equation}
where
\begin{equation}
		D= -\frac{\h^2}{2M}\frac{1}{n}\del\cdot(n\del)-i\frac{\h^2}{M}\del \theta\cdot\del=D_n+iD_\Theta
\end{equation}
and where $u_\vp=\vp$ and $v_\vp=\vp^*$ are considered independent.

The problem \eqref{eq:bogoliubov-matrix} is not hermitian, so that the eigenvalues are not in general real. However the matrix $\L$ is $\sigma_3$\textit{-pseudo-hermitian}, that is $\sigma_3\L^\dag\sigma_3=\L$. From this some properties follow. If $\e_i$ is an eigenvector, also $\e_i*$ must be one and modes with $\e_j=\e_i^*$ are called \textit{pseudo-degenerate}. Also, the non-positive definite inner product
\begin{equation}\label{eq:bogo-innerproduct}
	\braket{\vp|\psi}_B= \braket{\vp|\sigma_3|\psi}= \int d\vc{x}\left[u_\vp^*(\vc{x})u_\psi(\vc{x})-v_\vp^*(\vc{x})v_\psi(\vc{x})\right]
\end{equation}
is conserved in time. Finally, for any eigenvectors, $\p_i$ and $\p_j$
\begin{equation}
	\left(\e_i-\e_j^*\right)\braket{\vp_j|\vp_i}_B=0,
\end{equation}
so that modes with different eigenvalues are orthogonal and pseudo-degenerate modes have zero norm.

The energy of a mode is given by
\begin{equation}
	E_i=||\vp_i||_B\ \e_i
\end{equation}
so that, for positive frequencies, modes with a positive norm have a positive energy and ones with a negative norm have negative energy. For negative frequencies instead one will find the particle-hole symmetric modes of the ones at positive frequencies. Notice also that pseudo-degenerate modes have zero energy and can be thought as the simultaneous creation of pairs of modes with opposite energies.

We can distinguish two notions of instability: \textit{energetic instability} is the presence of negative energy modes (in the text negative norm modes at positive frequencies) and \textit{dynamical instability} is instead the presence of pseudo-degenerate modes, whose complex frequencies determines an exponential growth in time.

Dynamical instabilities can emerge, under a perturbation of the parameters of the system, when two modes of opposite-signed norm approach the same frequency. In fact while same-signed modes experience avoided crossing, eigenvalues of opposite-signed modes \textit{stick} together, that is the real part of the frequencies become degenerate while their imaginary parts acquire opposite values. This also happens in Hamiltonian systems,
where the norm is replaced by the Krein signature. For a proof of this band sticking in the Bogoliubov problem see for example \cite{lundh2006dynamic}.

Consider now the Klein--Gordon equation in the transverse flow metric \eqref{eq:transverse-metric}
\begin{equation}
	\del^2\p - \frac{1}{c^2}\left(\de_t+v_x\de_x\right)^2\p=0.
\end{equation}
It is useful to rewrite this equation as a system of first-order in time ones \cite{feshbach1958elementary} by defining
\begin{equation}
	\pi = -\frac{1}{c^2}(\de_t+v_x\de_x)\p,
\end{equation}
so that the equation takes the form
\begin{equation}\label{eq:kg-hamiltonian}
	\de_t\begin{pmatrix} \p \\ \pi \end{pmatrix} =
	\begin{bmatrix}
		-v_x\de_x & -c^2\\
		\\
		-\del^2 & -v_x\de_x
	\end{bmatrix}
	\begin{pmatrix} \p \\ \pi \end{pmatrix}.
\end{equation}
The associated conserved inner product is
\begin{equation}\label{eq:klein-innerproduct}
	\braket{\p_1|\p_2}_{KG}=i\int\d\vc{x}\ \left[\pi_1^*\p_2 - \p_1^*\pi_2\right].
\end{equation}

Let us compare this with another formulation of the Bogoliubov problem is in terms of the density and phase variations of the second equality in \eqref{eq:linearization}, that corresponds to the Bogoliubov-de Gennes equations \eqref{eq:bdg}. Define
\begin{equation}\label{eq:hydro-fields}
	\delta \tld{n} = \frac{M}{\h}\frac{\delta n}{n}=\frac{M}{\h}(\vp+\vp^*);\hspace{0.5cm} \delta\Theta=\frac{\vp-\vp^*}{2i},
\end{equation}
in terms of which the Bogoliubov problem \eqref{eq:linearization} takes the form
\begin{equation}\label{eq:bogo-densphase}
	\de_t\begin{pmatrix} \delta\Theta \\ \delta \tld{n} \end{pmatrix} =
	\begin{bmatrix}
		-\vc{v}_0\cdot\del & -\frac{\h^2}{4M^2}\frac{1}{n}\del(n\del)-c_s^2\\
		\\
		-\h^2\frac{1}{n}\del(n\del) & -\vc{v}_0\cdot\del
	\end{bmatrix}
	\begin{pmatrix} \delta\Theta \\ \delta \tld{n} \end{pmatrix},
\end{equation}
so that, for a constant density and for the desired velocity field, if we neglect the derivatives term with respect to $c_s^2$ in the second element of the first row we exactly obtain the Klein--Gordon problem \eqref{eq:kg-hamiltonian}. In terms of the fields \eqref{eq:hydro-fields} the Bogoliubov inner product \eqref{eq:bogo-innerproduct} becomes
\begin{equation}
	\braket{\varphi|\psi}_B=i\frac{\h}{M}\int d\vc{x}\left[\delta\tld{n}_\vp\delta\Theta_\psi - \delta\Theta_\vp\delta\tld{n}_\psi\right],
\end{equation}
that has the same shape of the Klein--Gordon inner product \eqref{eq:klein-innerproduct} so that all the considerations about norm and energy we made also hold true for the Klein--Gordon equation. The two degrees of freedom that emerge in BECs excitations from the mixing of positive and negative frequencies are here a consequence of the second order in time of the Klein--Gordon equation \cite{feshbach1958elementary}.

Finally, let us rewrite problem \eqref{eq:bogo-densphase} in terms of the fields $u_\vp$ and $v_\vp$ after taking the hydrodynamic approximation. This is the problem whose spectrum we calculated for the non-dispersive case in Section \ref{sec:instabilities}:
\begin{equation}\label{eq:bogo-hydrodynamic}
	i\h\de_t\begin{pmatrix} u_\vp\\ v_\vp \end{pmatrix} =
	\begin{bmatrix}
		\frac{D_n}{2}+iD_\theta+gn & -\frac{D_n}{2}+gn\\
		\frac{D_n}{2}-gn & -\left(\frac{D_n}{2}-iD_\theta\right)-gn
	\end{bmatrix}
	\begin{pmatrix} u_\vp\\ v_\vp \end{pmatrix}.
\end{equation}
Notice that with respect to the full dispersive problem \eqref{eq:bogoliubov-matrix} here the derivatives are distributed in all the matrix elements.

It is useful to make a last comment on the comparison of the \textit{dispersive} Bogoliubov problem and the \textit{dispersionless} Klein--Gordon one. The first one has a typical energy scale that is the one over which dispersive effects take place; this is fixed by the physics of the underlying condensate, that gives the natural units we used in our plots. The second one instead has no intrinsic physical scale and this is why in Figure \ref{fig:kg-dispersion} we rescaled the quantities using the frequency. Throughout the work, when we compare the two problems, we use instead the condensate natural units also for the problem in the hydrodynamic limit; this makes sense for the comparison but from the point of view of the Klein--Gordon problem we are introducing an arbitrary scale.

\bibliography{biblio.bib}

\end{document}